\documentclass{ar-1col-S2O}
\usepackage[numbers]{natbib}
\usepackage{url}

\usepackage{amssymb,amsmath}
\usepackage{multirow}
\usepackage[utf8]{inputenc}

\usepackage{caption}
\usepackage{subcaption}

\newcommand{\beq}{\begin{equation}}
\newcommand{\eeq}{\end{equation}}

\newcommand{\cO}{\mathcal{O}}
\newcommand{\cL}{\mathcal{L}}

\newcommand{\cM}{\mathcal{M}}

\newcommand{\abs}[1]{\left\lvert#1\right\rvert}

\newcommand{\bra}[1]{\langle #1|}
\newcommand{\ket}[1]{|#1\rangle}

\newcommand{\alphaNP}{\alpha_{\rm NP}}

\newcommand{\TeV}{\mathrm{TeV}}
\newcommand{\GeV}{\mathrm{GeV}}
\newcommand{\MeV}{\mathrm{MeV}}
\newcommand{\keV}{\mathrm{keV}}
\newcommand{\eV}{\mathrm{eV}}

\newcommand{\cm}{\mathrm{cm}}

\newcommand{\sigex}{\sigma_{\rm exp}}
\newcommand{\sigth}{\sigma_{\rm th}}

\newcommand{\drsq}{\ensuremath{\delta\langle r^2 \rangle}}

\newcommand{\mIS}[1]{\ensuremath{\overline{#1\vphantom{\bar{\nu}}}}}

\definecolor{red1}{cmyk}{0,1,1,0.3}

\jname{Xxxx. Xxx. Xxx. Xxx.}
\jvol{AA}
\jyear{YYYY}
\doi{10.1146/((please add article doi))}

\begin{document}

% Page header
\markboth{Delaunay--Karr--Soreq}{Atomic Spectroscopy Probes of New Physics}

% Title
\title{Atomic Spectroscopy Probes of New Physics}

%Authors, affiliations address.
\author{C\'edric Delaunay,$^1$ Jean-Philippe Karr,$^{2,3}$ and Yotam Soreq$^{4,5}$
\affil{$^1$Laboratoire d'Annecy de Physique Th\'eorique, CNRS -- USMB, 74940 Annecy, France,}
\affil{$^2$Laboratoire Kastler Brossel, Sorbonne Universit\'e, CNRS, ENS-Universit\'e PSL, Coll\`ege de France, 4 place Jussieu, F-75005 Paris, France}
\affil{$^3$Universit\'e Evry Paris-Saclay, Boulevard Fran\c cois Mitterrand, F-91000 Evry, France}
\affil{$^4$Physics Department, Technion – Israel Institute of Technology, Haifa 3200003, Israel}
\affil{$^5$Theoretical Physics Department, CERN, 1211 Geneva 23, Switzerland}
}

%Abstract
\begin{abstract}
Precision spectroscopy has long played a central role in testing the foundations of physics, from the early insights that led to the development of quantum mechanics to the validation of quantum electrodynamics and the determination of fundamental constants. 
Today, advances in atomic and molecular spectroscopy enable sensitive searches for physics beyond the Standard Model. 
A broad class of well-motivated extensions predicts new light degrees of freedom with feeble couplings to electrons, muons, and nucleons, giving rise to tiny spin-independent interactions that can be probed at low energies. 
In this review, we present a unified overview of spectroscopic searches for such interactions. 
We discuss the effective theoretical framework connecting fundamental interactions to atomic and nuclear observables, survey the key experimental and theoretical strategies, and review the atomic and molecular systems providing the strongest sensitivity. 
We conclude with updated spectroscopic constraints on representative benchmark models, highlighting the unique and complementary role of precision spectroscopy in exploring new fundamental interactions. 
\end{abstract}

%Keywords, etc.

\maketitle

%Table of Contents
\tableofcontents

%%%%%%%%%%%%%%%%%%%%%%%%%%%%%%%%%%%%%%%%%%%%%%%%%%%%%%%%%%%%%%%%
\section{Introduction}
%%%%%%%%%%%%%%%%%%%%%%%%%%%%%%%%%%%%%%%%%%%%%%%%%%%%%%%%%%%%%%%%

The Standard Model~(SM) of particle physics is the result of decades of remarkable theoretical and experimental progress and stands as one of the major achievements of modern physics. 
It provides an extraordinarily successful description of a wide range of phenomena across many orders of magnitude in energy and length scales, and has been directly tested down to distances of order $\sim10^{-19}$m.
From a theoretical perspective, the SM is internally consistent and could in principle remain valid up to very high energies, potentially as high as the Planck scale, $M_{\rm Pl}\sim10^{18}\,\GeV$. 

Despite these successes, the SM is known to be incomplete. 
It cannot account for neutrino oscillations, the observed matter–antimatter asymmetry of the Universe, or the existence of dark matter~(DM). 
These experimental facts, together with well-motivated theoretical considerations, provide compelling evidence for the existence of new physics beyond the Standard Model~(BSM). 
At present, however, there is no experimental indication of a preferred energy scale or a unique theoretical framework for such new physics; see, for example, Ref.~\cite{deBlas:2025gyz}.
One particularly well-motivated possibility is the existence of new light degrees of freedom with masses well below the GeV scale and feeble couplings to electrons, muons, and nucleons. 
Such states arise naturally in a variety of BSM scenarios, some of which are discussed below. 
In this regime, low-energy precision measurements, most notably atomic and molecular spectroscopy, offer powerful and often unique probes. 

In this review, we focus primarily on new spin-independent interactions and the spectroscopic techniques used to search for them. 
For broader discussions of tabletop experiments probing new physics, we refer the reader to Ref.~\cite{Safronova:2017xyt}, and for reviews dedicated to spin-dependent interactions, to Ref.~\cite{Cong:2024qly}.
Spectroscopic searches are complementary to probes based on astrophysical and cosmological observations, as discussed for example in Refs.~\cite{Caputo:2021eaa,Caputo:2024oqc,Caputo:2025avc}, as well as DM searches employing quantum sensing techniques, reviewed in Ref.~\cite{Antypas:2022asj}. 

This review is organized as follows. 
In Section~\ref{sec:LagToObs}, we discuss the matching from quark- and lepton-level interactions to effective descriptions at the nucleon and nuclear levels, for both spin-independent and spin-dependent interactions. 
We also review the resulting effective potentials and the associated energy shifts induced in atomic and molecular systems for the different models considered.  
In Section~\ref{sec:Tech}, we present the theoretical and experimental strategies employed in spectroscopic searches, along with the atomic and molecular systems most relevant for probing new forces. 
In particular, we distinguish between approaches based on direct comparisons between theory and experiment and those relying on symmetry arguments or special observables with reduced sensitivity to theoretical uncertainties. 
Section~\ref{sec:systems} provides a systematic review of the different spectroscopic systems suited to date, including few-electron atoms, many-electron atoms, exotic atoms and molecules. 
In Section~\ref{sec:codata-np}, we present updated combined constraints from spectroscopic data for several benchmark models inducing spin-independent interactions.  
We conclude with a summary and outlook in Section~\ref{sec:Summary}.

%%%%%%%%%%%%%%%%%%%%%%%%%%%%%%%%%%%%%%%%%%%%%%%%%%%%%%%%%%%%%%%%
\section{From Lagrangians to observables}
\label{sec:LagToObs}
%%%%%%%%%%%%%%%%%%%%%%%%%%%%%%%%%%%%%%%%%%%%%%%%%%%%%%%%%%%%%%%%

%%%%%%%%%%%%%%%%%%%%%%%%%%%%%%%%%%%%%%%%%%%%%%%%%%%%%%%%%%%%%%%%
\subsection{Generic Lagrangians}
%%%%%%%%%%%%%%%%%%%%%%%%%%%%%%%%%%%%%%%%%%%%%%%%%%%%%%%%%%%%%%%%

We focus on new spin-0 boson, $\phi$, and new spin-1 boson, $X_\mu$, collectively denoted as $\varphi=\phi,X_\mu$.
Here we give generic Lagrangians; see Section~\ref{sec:codata-np} for few model examples.
The relevant interaction Lagrangian at the GeV scale between a new spinless boson $\phi$ and SM fields is 
\begin{align}
    \label{eq:Lphiq}
    -\cL^{s=0}_{\rm int}
    =
    \frac{\phi}{f_\phi}\Bigg[
    &\sum_{\psi=\ell,q} m_\psi\bar\psi\left(\kappa_\psi+i\tilde{\kappa}_\psi\gamma_5\right)\psi
    \nonumber\\
    &+\frac{\alpha_s c_g}{4\pi}{\rm Tr}\,G_{\mu\nu}G^{\mu\nu}
    +\frac{\alpha_s \tilde{c}_g}{4\pi} {\rm Tr}\,G_{\mu\nu}\widetilde{G}^{\mu\nu}
    +\frac{\alpha c_\gamma}{4\pi} \,F_{\mu\nu}F^{\mu\nu}
    +\frac{\alpha \tilde{c}_\gamma}{4\pi} \,F_{\mu\nu}\widetilde{F}^{\mu\nu}\Bigg] \,,
\end{align}
where $\ell=e,\mu$ and $q=u,d,s$ denote charged leptons and quarks, respectively, with masses below $1\,\GeV$. 
The parameter $f_\phi$ represents the mass scale associated with $\phi$ interactions, such as the axion decay constant or the breaking of conformal symmetry for a dilaton. 
The gluon field strength is defined as 
$G_{\mu\nu}\equiv \partial_\mu G_\nu-\partial\nu G_\mu-ig_s[G_\mu,G_\nu]$, with $G_{\mu}\equiv G_{\mu}^a T^a$ and $T^a$ being the SU(3)$_c$ generators in the fundamental representation. 
The dual gluon field strength is $\widetilde{ G}^{\mu\nu}\equiv \epsilon^{\mu\nu\rho\sigma}G_{\rho\sigma}/2$, where $\epsilon^{\mu\nu\rho\sigma}$ is the Levi-Civita symbol with $\epsilon^{0123}=+1$. 
Similarly, $F_{\mu\nu}\equiv \partial_\mu A_\nu-\partial_\nu A_\mu$ is the photon field strength and $\widetilde{F}_{\mu\nu}$ is its dual.

In the limit where the couplings $\tilde{\kappa}_\psi$, $\tilde{c}_g$ and $\tilde{c}_\gamma$ vanish, $\phi$ is even under the product of charge conjugation~(C) and parity~(P) symmetries and behaves as a scalar. 
Conversely, if $\kappa_\psi=c_{g,\gamma}=0$, $\phi$ is CP-odd and behaves as a pseudoscalar. 
In the mixed case, $\phi$ has no definite CP quantum number, and its interactions violate CP symmetry.
The CP-odd coupling can be shifted between the $\phi G\widetilde{G}$ term and the $\phi \bar{q}\gamma_5q$ term through a chiral transformation, $q\to e^{ic\gamma_5}q$, where $c$ is a real constant. 
However, all physical quantities depend only on combinations of these couplings that are invariant under chiral transformations, as discussed in Refs.~\cite{Bauer:2021wjo,Bonnefoy:2022rik,Balkin:2025enj,Ovchynnikov:2025gpx}.\\

For a new spin-1 boson $X_\mu$, the relevant interaction Lagrangian is 
\begin{align}
    \label{eq:LXq}
    -\cL^{s=1}_{\rm int}
    = 
    X_\mu
    \sum_{\psi=\ell,q} \bar\psi\gamma^\mu
    \left(g^V_{\psi}+g^A_{\psi}\gamma_5\right)
    \psi\,.
\end{align}
We emphasize that $\cL^{s=1}_{\rm int}$ is expressed in the mass basis of $X_\mu$.
For example, a kinetically mixed dark photon, described by the interaction Lagrangian $-\frac{\varepsilon}{2}X_{\mu\nu}F^{\mu\nu}$~\cite{Holdom:1985ag,Okun:1982xi}, yields couplings of the form $g^V_\psi = e \varepsilon  q_\psi$ and $g^A_\psi=0$ after moving to the mass basis and canonical kinetic terms.

%%%%%%%%%%%%%%%%%%%%%%%%%%%%%%%%%%%%%%%%%%%%%%%%%%%%%%%%%%%%%%%%
\subsection{Matching to effective hadronic couplings}
%%%%%%%%%%%%%%%%%%%%%%%%%%%%%%%%%%%%%%%%%%%%%%%%%%%%%%%%%%%%%%%%

In atomic and molecular systems, the relevant degrees of freedom are nucleons,  nuclei, and, in some cases, charged pions and kaons. 
Our next step is to map the Lagrangian of Eqs.~\eqref{eq:Lphiq} and \eqref{eq:LXq} to the nucleon level.
The spin-0 hadronic-level interaction Lagrangian can be written as (see for example Refs.~\cite{Leutwyler:1989xj,Georgi:1986df})
\begin{align}
    \label{eq:Lphihad}
   - \cL^{s=0}_{\rm int, had}
    =
    \frac{\phi}{f_\phi}\Bigg[
    \sum_{N=n,p }m_N\bar{N}( g^S_N + ig^P_N\gamma_5)N
    +\sum_{\cM=\pi,K}g_\cM^S m_\cM^2 \cM^+\cM^- \Bigg] \, .
\end{align}
The $\phi$-nucleon couplings are given by
\begin{align}
    g^S_N 
    =& 
    \sum_q \kappa_q f^N_q  - \frac{ c_g}{9}\left(1-\sum_q f_q^N\right) \, , 
    \quad
    g^P_N 
    = 
    \sum_q \tilde{\kappa}_q \tilde{f}^N_q  + \tilde{c}_g \tilde{f}^N_g \, , 
\end{align}
where the coefficients $f^N_q$ and $\tilde{f}^N_q$ are provided in Table~\ref{tab:fhad}.
For the $\phi$ couplings to charged mesons, we have~\cite{Delaunay:2025lhl}
\begin{align}
    g^S_{\pi} 
    =&  
    \frac{\kappa_u+\kappa_d}{2}
    -\frac{c_g}{9}\left(1+\frac{m_\phi^2}{m_\pi^2}\right)
    \, ,\\
    g^S_{K}
    =& 
    \kappa_s + (\kappa_u-\kappa_s)\frac{m_\pi^2}{2m_K^2}
    -\frac{c_g}{9}\left(1+\frac{m_\phi^2}{m_\pi^2}\right)\,.
\end{align}
Couplings to neutral mesons are not considered in this analysis, as they do not form bound states with nuclei. Note that pseudoscalar couplings to mesons are absent due to CP symmetry. 
These couplings can appear only in CP-violating scenarios, where the mass state is a linear combination of CP-even and CP-odd states.

For spin-1, the hadronic-level interaction Lagrangian is 
\begin{align}
    \label{eq:Xhad}
    -\cL^{s=1}_{\rm int, had}
    =
    X_\mu \Bigg[\sum_{N=p,n}\bar{N}\gamma^\mu(g_N^V + g_N^A \gamma_5)N+\sum_\mathcal{M=\pi,K} ig_\mathcal{M}^V\left[\mathcal{M}^+(\partial^\mu\mathcal{M}^-)-(\partial^\mu\mathcal{M}^+)\mathcal{M}^-\right] \Bigg]\, .
\end{align}
The $X_\mu$-couplings are given by 
\begin{align}
    g^{V}_{p} =& 2g^{V}_u+g^{V}_d  \, , 
    \qquad\qquad\qquad\quad
    g^{V}_{n} =  g^{V}_u+2  g^{V}_d \, ,\\
    g^{A}_{p} =& h^p_u g^{A}_u+h^p_dg^{A}_d+h^p_sg^{A}_s  \, , 
    \qquad
    g^{A}_{n} =  h^n_u g^{A}_u+h^n_d g^{A}_d+h^n_s g^{A}_s \, ,
\end{align}
where the coefficients $h^N_{u,d}$ are given in Table~\ref{tab:fhad}, and the $X_\mu$-meson couplings are given by
\begin{align}
    g_\pi^V =2(g_d^V-g_u^V)\,, 
    \qquad 
    g_K^V = 2(g_s^V-g_u^V)\,.
\end{align}
Note that there are no axial couplings quadratic in the meson fields, as they violate parity. 
The leading axial interactions involve 3 meson fields and  are therefore irrelevant for spectroscopy.\\

%%%%%%%%%%%%%%%%%%%%%%
\begin{table}[t]
\begin{center}
\begin{tabular}{ccccccccccc}
\hline\hline
$N$ & $f_u^N$  & $f_d^N$  & $f_s^N$  &  
$\tilde{f}_u^N$  & $\tilde{f}_d^N$  & $\tilde{f}_s^N$ & $\tilde{f}^N_g$ 
 & $h_u^N$ & $h_d^N$ & $h_s^N$\\
\hline
$n$& 0.015 & 0.037 & 0.046 & -0.92 & 0.43  & -0.35 & -0.24 & -0.38 & 0.90 & -0.031\\
$p$& 0.017 & 0.033 & 0.046 & 0.43 & -0.92 & -0.35 & -0.24 & 0.90 & -0.38 & -0.031\\
\hline\hline
\end{tabular}
\end{center}
\caption{Form factors for scalar, pseudoscalar and axial-vector couplings of $\varphi$ to nucleons in the zero momentum limit, taken from Refs.~\cite{Bishara:2017pfq,Haxton:2024lyc}.}
\label{tab:fhad}
\end{table}
%%%%%%%%%%%%%%%%%%%%%%

%%%%%%%%%%%%%%%%%%%%%%%%%%%%%%%%%%%%%%%%%%%%%%%%%%%%%%%%%%%%%%%%
\subsection{Mapping to non-relativistic potentials}
%%%%%%%%%%%%%%%%%%%%%%%%%%%%%%%%%%%%%%%%%%%%%%%%%%%%%%%%%%%%%%%%

The most relevant observables in atomic and molecular systems are transition frequencies, which can be measured with high precision. 
The presence of a new force carriers $\varphi$ induces small shifts in the spectroscopic lines, which can be described by an effective potential $V_\varphi(\boldsymbol{r})$ between two particles $1$ and $2$ separated by a distance $\boldsymbol{r}$ (bold symbols denote spatial vectors).  
Following the non-relativistic derivation in Refs.~\cite{Fadeev:2018rfl,Frugiuele:2021bic,Cong:2024qly}, we neglect fine-structure effects arising from spin-orbit couplings, as these typically constitute sub-leading contributions from new physics~\cite{Cong:2024qly}. 
Furthermore, to accommodate situations involving either a single nucleus or two nuclei, we distinguish explicitly between spin-independent~($g_{\rm SI}$) and spin-dependent~($g_{\rm SD}$) couplings in the effective non-relativistic potentials.

The $\phi$-induced potential is given by
\begin{align}
    \label{eq:Vphi}
    V_\phi(\boldsymbol{r}) 
    = 
    V_{SS}(\boldsymbol{r}) 
    + V_{PS}(\boldsymbol{r}) 
    + V_{PP}(\boldsymbol{r}) \, ,
\end{align}
where 
\begin{align}
    \label{eq:VSS}
    V_{SS}(\boldsymbol{r})
    =&
    -g^S_{{\rm SI},1} g^S_{{\rm SI},2}\,\frac{e^{-m_\phi r}}{4\pi r} \, , \\
    \label{eq:VPS}
    V_{PS}(\boldsymbol{r})
    =&
    -g^P_{{\rm SD},1} g^S_{{\rm SI},2}\, \boldsymbol{\sigma}_1 \cdot \boldsymbol{\hat{r}} 
    \left( \frac{1}{r}+ m_\phi \right)
    \frac{e^{-m_\phi r}}{8\pi r} +1\leftrightarrow2\, , \\
    \label{eq:VPP}
    V_{PP}(\boldsymbol{r})
    =&
    -g^P_{{\rm SD},1}g^P_{{\rm SD},2}
    \Bigg[ \boldsymbol{\sigma}_1 \cdot \boldsymbol{\sigma}_2\left(\frac{1}{r^2}+\frac{m_\phi}{r}+
     \frac{4\pi r}{3}\delta^3(\boldsymbol{r}) \right) \nonumber\\
    & \qquad\qquad\qquad 
    -3(\boldsymbol{\sigma}_1\cdot \boldsymbol{\hat{r}})(\boldsymbol{\sigma}_2\cdot \boldsymbol{\hat{r}})
    \left( \frac{1}{r^2}+\frac{m_\phi}{r}+\frac{m^2_\phi}{3} \right)
    \Bigg] \frac{e^{-m_\phi r}}{16\pi m_1m_2 r} \, ,    
\end{align}
are the scalar-scalar, pseudoscalar-scalar and pseudoscalar-pseudoscalar contributions, respectively.
Here $\boldsymbol{\sigma}_{1,2}$ denotes the spin of particles $1$ and $2$, $r\equiv |\boldsymbol{r}|$ and $\boldsymbol{\hat r}\equiv \boldsymbol{r}/r$.  The $X_\mu$ induced potential is given by
\begin{align}
    \label{eq:VX}
    V_X(\boldsymbol{r}) = V_{VV}(\boldsymbol{r}) + V_{AV}(\boldsymbol{r}) + V_{AA}(\boldsymbol{r}) \, ,
\end{align}
where 
\begin{align}
    \label{eq:VVV}
    V_{VV}(\boldsymbol{r})
    =&
   g^V_{{\rm SI},1} g^V_{{\rm SI},2}\frac{e^{-m_X r}}{4\pi r}
    +
    g^V_{{\rm SD},1} g^V_{{\rm SD},2}
    \Bigg[ \boldsymbol{\sigma}_1 \cdot \boldsymbol{\sigma}_2\left(
    \frac{1}{r^2}+ \frac{m_X}{r} -\frac{8\pi r}{3}\delta^3(\boldsymbol{r})+m_X^2 
    \right) \nonumber\\
    & \quad\quad 
    -3(\boldsymbol{\sigma}_1\cdot \boldsymbol{\hat{r}})(\boldsymbol{\sigma}_2\cdot \boldsymbol{\hat{r}})
    \left( \frac{1}{r^2}+\frac{m_X}{r}+\frac{m^2_X}{3} \right)
    \Bigg]
    \frac{e^{-m_X r}}{16\pi m_1m_2r} \, ,  \\
    \label{eq:VAV}
    V_{AV}(\boldsymbol{r})
    =&
    g^A_{{\rm SD},1} g^V_{{\rm SI},2}\,
    \boldsymbol{\sigma}_1\cdot \left\{ \frac{\boldsymbol{p}_1}{m_1} \!- \! \frac{\boldsymbol{p}_2}{m_2},
    \frac{e^{-m_X r}}{8\pi r} \right\} \nonumber\\
    &\qquad\qquad
    - g_{{\rm SD},1}^Ag_{{\rm SD},2}^V(\boldsymbol{\sigma}_1\times \boldsymbol{\sigma}_2)\cdot \boldsymbol{\hat{r}}
    \left( \frac{1}{r}+m_X \right)\frac{e^{-m_Xr}}{8\pi m_2 r}
 +1\leftrightarrow 2\, , \\
    \label{eq:VAA}
    V_{AA}(\boldsymbol{r})
    =&
    -g^A_{{\rm SD},1} g^A_{{\rm SD},2}
    \Bigg[ \boldsymbol{\sigma}_1\cdot\boldsymbol{\sigma}_2\left(\frac{1}{r^2} + \frac{m_X}{r }+  \frac{4\pi r}{3}\delta^3(\boldsymbol{r})+ m_X^2 \right) \nonumber\\
  & \qquad\qquad\qquad
  -3(\boldsymbol{\sigma}_1\cdot\boldsymbol{\hat{r}})(\boldsymbol{\sigma}_2\cdot\boldsymbol{\hat{r}})
    \left(\frac{1}{r^2}+\frac{m_X}{r}+\frac{m_X^2}{3} \right)
    \Bigg]\frac{e^{-m_Xr}}{4\pi m_X^2r} \, ,
\end{align}
where $\{\cdot,\cdot\}$ is the anti-commutator. 

SI interactions arise from the exchange of scalar and vector particles. 
For elementary leptons, the corresponding couplings are $g_{{\rm SI},\ell}^{S}=m_\ell\kappa_\ell/f_\phi$ and $g_{{\rm SI},\ell}^V=g_\ell^V$, while for nucleons $g_{{\rm SI},N}^{S/V}=g^{S/V}_N$. 
For a nucleus with atomic number $Z$ and mass number $A$, the effective SI couplings are given by 
\begin{align}
    g^{S/V}_{\rm SI} 
    = 
    (A-Z)g^{S/V}_n+ Z g^{S/V}_p \, .
\end{align}
SD interactions are induced by pseudoscalar, axial-vector and vector particles. 
For leptons, the corresponding couplings are $g_{{\rm SD},\ell}^P=m_\ell\tilde\kappa_\ell/f_\phi$ and $g_{{\rm SD},\ell}^{A/V}=g_\ell^{A/V}$, while for nucleons $g_{{\rm SD},N}^{P/A/V}=g_N^{P/A/V}$. 
The effective nuclear couplings can be approximated by using the Schmidt model or the Flambaum-Tedesco model, as discussed in Refs.~\cite{JacksonKimball:2014vsz,Flambaum:2006ip,Stadnik:2014xja}, and are given by
\begin{align}
    g^{P/A/V}_{\rm SD}
    =
    b_n g^{P/A/V}_n + b_p g^{P/A/V}_p \, ,
\end{align}
where $b_{n,p}$ denote the fractions of the nuclear spin carried by neutrons and protons, respectively, and thus depend on the specific nucleus. 
A dedicated review on SD interactions can be found in Ref.~\cite{Cong:2024qly}.

It is worth noting that nuclear SI couplings scale linearly with the number of nucleons, $g_{\rm SI}^{S/V}\sim\cO(A)$, whereas SD couplings do not, $g^{P/A/V}_{\rm SD} \sim \cO(1)$. 
The potentials $V_{SS}$, $V_{PP}$, $V_{VV}$ and $V_{AA}$ preserve C, P and time-reversal (T) symmetries, while  
$V_{PS}$ violates CP and T symmetries. 
The $V_{AV}$ potential violates parity but conserves T and CP. 
These potentials are derived in the non-relativistic limit and are therefore  valid only for mediator masses $m_\varphi \lesssim \sqrt{m_1m_2}$, see for example Ref.~\cite{Delaunay:2021uph}.\\

Finally, in the above discussion we have considered only tree-level exchange of the new particle.
However, it is possible that the leading contribution arises instead at the one-loop level. 
A well-known example is the force induced by the exchange of a pair of neutrinos~\cite{Feinberg:1968zz}. 
Another possibility is that new light particles generate such an effect at loop level (see for example Ref.~\cite{Bauer:2023czj}).

%%%%%%%%%%%%%%%%%%%%%%%%%%%%%%%%%%%%%%%%%%%%%%%%%%%%%%%%%%%%%%%%
\subsection{Energy shifts from new physics}
%%%%%%%%%%%%%%%%%%%%%%%%%%%%%%%%%%%%%%%%%%%%%%%%%%%%%%%%%%%%%%%%

While state-of-the-art calculations are required for the SM contributions, the effects of new physics can typically be evaluated at first order in perturbation theory.
The physical systems of interest can be broadly divided into two categories.
The first consists of systems in which the wave functions of the interacting particles overlap and extend throughout the entire volume of the system, such as electrons bound in atoms.
The second comprises systems in which the wave functions of the interacting particles are well separated, as in nucleus-nucleus interactions in molecules.\\ 

To illustrate the general features of the first case, we consider a non-relativistic hydrogen-like system. 
A state $\ket{\Psi_{nl}}$, characterized by the principal quantum number $n$ and the angular momentum quantum number $l$, has a radial wave function of the form
\begin{align}
    \Psi_{nl}(r)
    = 
    \frac{e^{-r/a_n}}{a_n^{3/2}} \sum_{l'=l}^{n-1}c_{l'}
    \left(\frac{r}{a_n}\right)^{l'}    
\end{align}
where $a_n$ denotes the characteristic size of the state, for instance, the Bohr radius for the ground state, and the $c_{l}$'s are known coefficients. 
Since all effective potentials considered above scale as $g_1g_2e^{-m_\varphi r}/r$, the correction to the energy level $E_{nl}$ induced by the exchange of $\varphi$ can be estimated as
\begin{align}
    \label{eq:Evarphin}
    E^\varphi_{nl}
    =
    \bra{\Psi_{nl}}V_\varphi \ket{\Psi_{nl}}
    =
    \frac{g_1 g_2}{a_n} \frac{f_{nl}(m_\varphi a_n)}{(2+ m_\varphi a_n)^{2(1+l)}} \, ,
\end{align}
where $f_{nl}(m_\varphi a_n)$ is a state-dependent function that encodes  the detailed structure of the wave function and the potential. 
For example, in the case of a Yukawa potential $V_{SS}(r)$ and $l=0$, this function reduces to a constant. 
More generally, Eq.~\eqref{eq:Evarphin} implies that for very light mediators, $m_\varphi a_n \ll 1$, the energy shift scales as
\begin{align}
    E_{nl}^\varphi 
    \propto 
    \frac{g_1 g_2}{a_n}f_{nl}(0)\,,
\end{align}
whereas for heavy mediators, $m_\varphi a_n \gg 1$, it behaves as
\begin{align}
    E_{nl}^\varphi 
    \propto 
    \frac{g_1 g_2}{a_n}
    \frac{f_{nl}(\infty)}{(a_n m_\varphi)^{2(l+1)}}\,. 
\end{align}

The second case, involving well-separated wave functions, is conceptually simpler.
There, the energy correction can be well approximated by evaluating the potential at the equilibrium separation,  
\begin{align}
    \label{eq:Evarphin2}
    E^\varphi_{nl}
    \simeq
    V_\varphi(r=r_{\rm eq}) \propto g_1g_2  \frac{e^{-m_\varphi r_{\rm eq}}}{r_{\rm eq}} \, ,
\end{align}
where $r_{\rm eq}$ is the average distance between the interacting particles. 
This situation arises, for instance, in nucleus-nucleus interactions within molecules.
Since in many cases $r_{\rm eq}\sim a_0$, the parametric dependence of the new physics contribution is similar to that in Eq.~\eqref{eq:Evarphin}. 
However, for mediator masses $m_\varphi \gtrsim r_{\rm eq} $, the BSM effect becomes exponentially suppressed, in contrast to the first case, where the decoupling proceeds as a power-law $\propto 1/m^{2(l+1)}_\varphi$.

The new-physics contribution is approximately constant for mediator masses $m_\varphi\lesssim1/a_n$, except in cases where degeneracies occur. 
Consequently, the limit $m_\varphi\to0$ provides a good approximation to the energy shift from new physics up to masses $m_\varphi\sim1/a_n$, beyond which the interaction begins to decouple.
The decoupling behavior depends on the spatial structure of the system. 
It scales as $m_\varphi^{2(l+1)}$ when the wave function fills the entire volume, and becomes exponential when the two interacting particles are well separated.\\

For several of the potentials discussed above, compact expressions can be obtained in the $m_\varphi\to0$ limit for a hydrogen-like atom with nuclear charge $Z$ and Bohr radius defined as $a_0 \equiv  (\alpha m_r)^{-1}$, where $m_r= m_1m_2/(m_1+m_2)$ is the reduced mass of the two interacting particles. 
The atomic states $\ket{nljFM_F}$ are labelled by the principal quantum number $n$, the quantum number $j$ of the total angular momentum of the bound particle $\boldsymbol{J} = \boldsymbol{L} + \boldsymbol{S}$, with $\boldsymbol{L}$ and $\boldsymbol{S}$ denoting its orbital and spin angular momenta, respectively, and the quantum number $F$ of total atomic angular momentum $\boldsymbol{F} = \boldsymbol{J} + \boldsymbol{I}$, where $\boldsymbol{I}$ is the nuclear spin. 
The quantum number $M_F$ denotes the projection of $\boldsymbol{F}$ along, for instance, the direction of an external magnetic field. 

As an illustration, we consider the case of a lepton with spin $s=1/2$ bound to a nucleus of spin $I$. 
The expectation values of the relevant potentials for spin-0 and spin-1 mediators are given by
\begin{align}
    \bra{nljFM_F}V_{SS}\ket{nljFM_F} 
    =& 
    -\frac{g_{\rm SI,1}^Sg_{\rm SI,2}^S}{4\pi}\frac{Z}{a_0n^2 }\, , \\
    \bra{nljFM_F}V_{PP}\ket{nljFM_F}
    =&
    -\frac{g^P_{{\rm SD},1}g^P_{{\rm SD},2}}{16\pi m_1 m_2} \frac{Z^3}{a_0^3n^3}\mathcal{C}_{ljIF}\,, \\
     \bra{nljFM_F} V_{VV} \ket{nljFM_F} 
    =&
    \frac{g^V_{{\rm SI},1} g^V_{{\rm SI},2}}{4\pi} \frac{Z}{a_0n^2}
    +
    \frac{g^V_{{\rm SD},1} g^V_{{\rm SD},2}}{16\pi m_1 m_2} \frac{Z^3}{a_0^3n^3} \mathcal{C}_{ljIF}\,,
\end{align}
where we defined
\begin{align}
    \mathcal{C}_{ljIF}
    \equiv& 
    \frac{\left[ j(j+1) \!-\! 3/4 \!-\! l(l+1)\right] \left[F(F+1) \!-\! j(j+1) \!-\! I(I+1)\right]}
    {l(l+1/2)(l+1)j(j+1)}
\end{align}
for $l\geq 1$, and $\mathcal{C}_{0 s IF}\equiv (8/3)[F(F+1)-2I(I+1)]$ for $l=0$. 
The $V_{PS}$ and $V_{AV}$ violate parity and CP symmetries, respectively. Consequently, their expectation values vanish unless one considers observables that are explicitly sensitive to P- or CP-violating effects. 
Finally, the axial-axial potential $V_{AA}$ diverges in the limit $m_X\to0$, indicating that this regime cannot be treated consistently  within the nonrelativistic theory. 
This reflects the fact that the axial current is generally anomalous, and the consistency of the underlying theory requires anomaly cancellation. 

%%%%%%%%%%%%%%%%%%%%%%%%%%%%%%%%%%%%%%%%%%%%%%%%%%%%%%%%%%%%%%%%
\section{Search strategies}
\label{sec:Tech}
%%%%%%%%%%%%%%%%%%%%%%%%%%%%%%%%%%%%%%%%%%%%%%%%%%%%%%%%%%%%%%%%

In this section, we review the basic principles underlying various spectroscopy-based methods for probing new physics beyond the SM.
Broadly, these methods can be classified into two categories, depending on the relative size of the theoretical (from SM calculations) and experimental uncertainties, denoted by $\sigth$ and $\sigex$, respectively.
\begin{enumerate}
\item \textit{Direct theory-experiment comparison.} 
This category includes systems for which the SM theoretical predictions are under good control, typically satisfying $\sigth \lesssim\sigex$. 
In such cases, a detailed comparison between theory and experiment provides a powerful probe of new physics. 
\item \textit{Theory-independent observable comparison.} 
This category comprises systems for which the theoretical uncertainty is large or poorly controlled, $\sigth\gg \sigex$, rendering a direct comparison between theory and experiment ineffective.
Instead, sensitivity to new physics is achieved by exploiting special properties of the system, such as symmetries (often discrete), factorization between different physical scales or characteristic time dependences, that are respected by the SM. 
Since no detailed theoretical calculation is required in this approach, new physics that violates these properties can be  probed efficiently.
\end{enumerate}
Below, we discuss these two categories in more details and provide representative examples. 

%%%%%%%%%%%%%%%%%%%%%%%%%%%%%%%%%%%%%%%%%%%%%%%%%%%%%%%%%%%%%%%%
\subsection{Direct theory-experiment comparison} 
\label{sec:th-exp}
%%%%%%%%%%%%%%%%%%%%%%%%%%%%%%%%%%%%%%%%%%%%%%%%%%%%%%%%%%%%%%%%

Standard Model predictions of atomic and molecular structure are based on the theory of quantum electrodynamics~(QED). 
This is also the case for properties of elementary particles such as the anomalous magnetic moments of the electron and the muon~\cite{Volkov:2024yzc,Aoyama:2024aly,Aliberti:2025beg}, $a_{e,\mu}\equiv(g-2)_{e,\mu}/2$, which will not be discussed in this section. 
These predictions depend on the values of several fundamental constants~\cite{Mohr:2024kco}, including the fine-structure constant, $\alpha$, the masses of the electron, proton, and light nuclei, as well as nuclear charge radii. 

QED calculations follow a perturbative approach in which corrections are expressed as an expansion in $\alpha$, starting from the bound-state wave functions. 
Since QED is intrinsically a relativistic theory, its most direct formulation is based on the Dirac equation. 
However, the Dirac equation is a one-body equation, and a rigorous treatment of bound states involving two interacting particles requires the Bethe-Salpeter formalism~\cite{Salpeter:1951sz}. 
While such an approach exists for two-body systems, its extension to systems with more interacting particles is prohibitively complex. 

This difficulty motivated the development of nonrelativistic QED~(NRQED) methods~\cite{Caswell:1985ui,Kinoshita:1995mt,Pachucki:2005,Jentschura:2022xuc}, which rely on an effective nonrelativistic Lagrangian describing the interaction of charged particles with the electromagnetic field. 
In this framework, bound-state wave functions are obtained by solving the few-body Schr\"odinger equation, and QED corrections are then computed using nonrelativistic perturbation theory. 
NRQED is the most widely used approach for weakly bound systems ($Z<10$) containing three or more particles, while studies of weakly bound two-body systems typically combine relativistic and nonrelativistic techniques~\cite{Eides:2007exa,Yerokhin:2018gna,Mohr:2024kco}.

Both the numerical resolution of the few-body Schr\"odinger equation and the derivation of QED corrections become increasingly challenging as the number of particles grows. 
The complexity of QED calculations is particularly sensitive to the number of {\em dynamical} particles, such as electrons or other orbiting particles, as well as nuclear recoil effects. 
Interaction between a light particle and a heavy nucleus can often be treated by considering the light particle moving in an external nuclear potential, with recoil corrections added perturbatively as an expansion in the mass ratio.  
No such simplification exists for interaction between light particles, such as electron-electron interactions in helium or electron-positron interactions in positronium. 
As a result, theoretical precision decreases rapidly with the number of light particles, and in practice the most stringent constraints on new physics forces are obtained from spectroscopy of systems containing one or two leptons.\\ 

Systems of interest for a direct comparison between theory and experiment include the following:
\begin{itemize}
\item \textit{Hydrogen-like atoms}, which are sensitive to electron-proton and electron-neutron couplings;
\item \textit{One-electron molecules}, such as molecular hydrogen ions, which probe proton-proton, proton-neutron, electron-proton and electron-neutron couplings;
\item \textit{Muonic hydrogen-like atoms}, sensitive to muon-proton and muon-neutron couplings;
\item \textit{Muonium}, which probes electron-antimuon couplings;
\item \textit{Positronium}, sensitive to electron-positron couplings;
\item \textit{Helium-like atoms}, probing electron-electron, electron-proton and electron-neutron couplings;
\item \textit{Muonic helium-like ions}, such as $\mu$He$^+$ probing muon-proton and muon-neutron couplings;
\item \textit{Two-electron molecules}, such as molecular hydrogen, which are sensitive to electron-electron, electron-proton, electron-neutron, proton-proton, proton-neutron and neutron-neutron couplings;
\item \textit{Exotic atoms}, in which electrons are replaced by hadrons, typically antiprotons. Examples include antiprotonic helium and antiprotonic heavy atoms such as lead, neon or xenon, which are sensitive to proton-antiproton and neutron-antiproton couplings.
\end{itemize}

The restriction to few-body systems is not absolute. 
In some cases, many-body atomic systems can be treated to very good approximation as effective two-body systems, because one particle orbits the nucleus at a distance that is either much smaller or much larger than that of the remaining electronic cloud. 
This separation of scales allows for accurate SM predictions. A couple of examples will be discussed in Section~\ref{sec:systems}.

Theoretical precision can also be improved by considering specific combinations of energy levels designed to eliminate a large fraction of QED contributions that depend on contact interactions, namely on the squared wave function at zero separation between two particles.
A generic example of such a combination is
\begin{align}
    \Delta_{ab} 
    = 
    \frac{E_a}{\langle \delta(\boldsymbol{r}) \rangle_a} 
    - \frac{E_b}{\langle \delta(\boldsymbol{r}) \rangle_b} \,,
\end{align}
where $E_{a,b}$ denote the energies of two different levels of the system, $\langle \delta(\boldsymbol{r})\rangle_{a,b}$ are their contact densities, and $\boldsymbol{r}$ is the relative position of the two particles. 
Similar cancellations can be achieved by studying isotope shifts of electronic transitions, where a large part of the theoretical contributions cancels between isotopes, as in hydrogen-deuterium or $^3$He-$^4$He isotope shifts~\cite{Pachucki:2017xcg,Delaunay:2017dku}.

A major limitation to theoretical precision arises from uncertainties in nuclear structure, such as nuclear charge radii and nuclear polarizabilities. 
These effects are particularly important for hyperfine structure and for muonic atoms, where the muon's orbital radius is smaller by a factor of approximately $m_{\mu} / m_e \sim 200$ relative to ordinary atoms, leading to a strong enhancement of nuclear effects. 
The specific differences described above can also be used to suppress nuclear contributions.
Alternatively, nuclear parameters can be extracted independently, for example from scattering experiments or from a limited set of dedicated spectroscopic measurements.\\ 

Finally, it is important to note that transition frequencies depend on fundamental constants whose values may themselves be determined using some of the spectroscopic measurements employed to constrain physics beyond the SM. 
This introduces a potential loophole that requires careful treatment.
Ideally, constraints should be derived from combined analyses that include all relevant measurements used to determine these constants~\cite{Jaeckel:2010xx,Karshenboim:2010ck,Jones:2019qny,Delaunay:2022grr,Potvliege:2024xly}. 
Such a global analysis will be discussed in more detail in Section~\ref{sec:codata-np}. 
Reduced analyses based on subsets of data are also possible, provided correlations are sufficiently weak, and can be useful for isolating the contributions of individual systems; see for example Refs.~\cite{Jaeckel:2010xx,Jones:2019qny,Alighanbari:2025}. 

\vspace{3mm}

%%%%%%%%%%%%%%%%%%%%%%%%%%%%%%%%%%%%%%%%%%%%%%%%%%%%%%%%%%%%%%%%
\subsection{Theory-independent observable comparison}
\label{sec:TheoryFree}
%%%%%%%%%%%%%%%%%%%%%%%%%%%%%%%%%%%%%%%%%%%%%%%%%%%%%%%%%%%%%%%%

In many systems, SM theoretical predictions are far less precise than the corresponding experimental measurements.
For instance, atomic clocks have reached accuracies at the $10^{-19}$ level~\cite{Arnold:2025opticalclocks,Marshall:2025,Aeppli2024}, whereas the corresponding theoretical uncertainties are much larger.
In such cases, a direct comparison between SM predictions and experimental data does not provide optimal sensitivity to new physics.

To overcome this limitation, one can focus on observables that do not rely on precise theoretical predictions, but instead exploit robust properties of the SM that are expected to hold to very high accuracy and can be significantly altered by new physics.  
Several complementary strategies fall into this category. 
First, one can construct observables based on discrete symmetries, most notably P, CP and T symmetries, which are conserved by QED and QCD (up to a possible $\theta_{\rm QCD}$ term that is severely constrained~\cite{Abel:2020pzs}) and only weakly violated in the SM by the weak interaction.  
Second, one can search for time-dependent signals, as expected for example from oscillating or transient backgrounds associated with DM. 
Third, one can use data-driven methods, in which combinations of experimental observables are formed so that dominant theoretical uncertainties cancel.
Importantly, the interpretation of such observables relies primarily on symmetry arguments rather than on high-precision theoretical calculations, making them especially powerful in regimes where theoretical uncertainties are large.

%%%%%%%%%%%%%%%%%%%%%%%%%%%%%%%%%%%%%%%%%%%%%%%%%%%%%%%%%%%%%%%%
\subsubsection{Discrete symmetries: parity violation}
%%%%%%%%%%%%%%%%%%%%%%%%%%%%%%%%%%%%%%%%%%%%%%%%%%%%%%%%%%%%%%%%

Parity violation (PV) induced by the SM weak interaction is well established across a wide range of energy and distance scales. 
Its first observation occurred in nuclear $\beta$-decays, following the seminal work of Lee and Yang~\cite{Lee:1956qn} and the experimental confirmation by Wu and collaborators~\cite{Wu:1957my}, and it has since been extensively studied in high-energy collider experiments~\cite{ParticleDataGroup:2024cfk}.
At atomic scales, PV is firmly established through measurements of atomic parity violation~(APV)~\cite{Bouchiat:1974kt,Wood:1997zq}.
Future experiments on chiral molecules, in which the spectra of two enantiomers (also called optical isomers, which are mirror images of each other) are compared, provide an additional sensitive probe of PV~\cite{Letokhov:1975zz,PhysRevLett.83.1554}. 
Both APV and chiral-molecule experiments are sensitive to PV effects from the SM weak interaction as well as from new physics interactions~\cite{Ramsey-Musolf:1999qyv,Bouchiat:2004sp,Baruch:2024fbh,Gaul:2020bdq}.
At the operator level, PV interactions typically involve axial-vector couplings and generate effective potentials proportional to spin-momentum structures such as $\boldsymbol{\sigma}\cdot \boldsymbol{p}$ and or two-body terms like $(\boldsymbol{\sigma}_1\times\boldsymbol{\sigma}_2)\cdot \hat{\boldsymbol{r}}$. 
These characteristic signatures allow PV observables to be isolated experimentally through symmetry-based selection rules, largely independent of detailed atomic-structure calculations. 
While APV primarily probes electron-nucleus couplings, chiral molecules can be sensitive to both electron-nucleus and nucleus-nucleus couplings. 
PV beyond the SM typically arises from the exchange of a new spin-1 mediator $X_\mu$ with both vector and axial couplings, described by interactions of the form $\bar{\psi}X_\mu\gamma^\mu\psi$ and $\bar{\psi}X_\mu\gamma^\mu\gamma^5\psi$. 
Observable PV effects are proportional to the product of vector and axial couplings, $g^A g^V$, and are conveniently encoded in the effective potential $V_{AV}$.\\

In APV experiments, one considers atomic transitions that are forbidden by parity selection rules respected by QED and QCD~\cite{Bouchiat:1974kt}, but become allowed through PV interactions. 
A canonical example is the  $6S-7S$ transition in caesium~\cite{Wood:1997zq}. 
PV interactions induce a small mixing between states of opposite parity, such as $S$ and $P$ states, characterized by a mixing amplitude $A_{\rm PV}$.
The corresponding transition rate scales as $\abs{A_{\rm PV}}^2$, which is too small to be observed directly. 
To enhance the effect, an external electric field is applied, inducing a parity-conserving amplitude $A_{\rm E}$ which interferes with the PV amplitude. 
The total transition rate is then proportional to 
\begin{align}
    \label{eq:APVrate}
    \abs{A_{\rm E} + A_{\rm PV}}^2
    \simeq
    \abs{A_{\rm E}}^2+ 2{\rm Re}(A_{\rm E}A_{\rm PV})\, ,  
\end{align} 
allowing the linear PV contribution to be isolated by exploiting angular-momentum selection rules that control the relative sign of the interference term. 
An enhanced sensitivity can be achieved in the dysprosium atom by using two nearly degenerate states with opposite parity and an electric-dipole transition between them~\cite{PhysRevA.50.132,Leefer:2014tga}, yielding a large $A_{\rm E}$.

APV probes electron-nucleus interactions in which the electronic current is parity odd, $\bar{e}\gamma^\mu\gamma_5e$, while the nuclear current is parity even, $\bar{N}\gamma^\mu N$.
This structure arises because P-odd effects require a spin flip, which scales as $\boldsymbol{p}/m$ and is therefore much larger for electrons than for nucleons. 
Both SM and BSM interactions contribute to APV.
In the SM, the effect is mediated at energies below the $W$-boson mass by dimension-six four-fermion operators of the form  $(\bar{e}\gamma_\mu\gamma_5 e)( \bar{q}\gamma^\mu q)$, suppressed by the Fermi constant $G_F\approx1.166\times 10^{-5}\,\GeV^{-2}$. 
BSM contributions are described by the potential $V_{AV}$, involving a new spin-1 state with axial couplings to electrons and vector couplings to nucleons, and are therefore sensitive to the product $g^A_e \times g^V_N$.

A key difference between SM and BSM contributions lies in the mediator mass. 
In the SM, the weak interaction can be treated as a contact interaction at atomic scales, whereas a BSM mediator may be much lighter. 
The typical momentum transfer in APV is of order MeV~\cite{Bouchiat:2004sp}.
For mediator masses well above this scale, the BSM contribution scales as $g^A_eg^V_N/m_X^2$ and effectively appears as a correction to the weak charge of the nucleus~\cite{Ramsey-Musolf:1999qyv}.
For masses $m_X\lesssim 10\,\MeV$, the interaction becomes long-ranged and exhibits only a mild dependence on $m_X$~\cite{Bouchiat:2004sp}.\\ 

Chiral molecules offer a complementary probe of PV. 
Their left- and right-handed enantiomers are related by parity transformation, so any spectral difference between them constitutes a direct signal of PV~\cite{Letokhov:1975zz}. 
In the SM, the weak interaction induces energy differences at the level of order Hz in vibrational modes; see for example Refs.~\cite{PhysRevLett.84.3807,PhysRevA.105.012820}. 
The sensitivity of chiral molecules to new PV forces has been explored in recent studies~\cite{Baruch:2024fbh}, showing that new PV electron-nucleon interactions could modify vibrational spectra and be experimentally
probed. 
The resulting BSM sensitivity is comparable to that of APV measurements and subject to similar theoretical limitations. 
In addition, measurements of hyperfine structure in rotational transitions, where SM weak effects are negligible, can probe new long-range PV nucleon-nucleon interactions. 
In this context, the molecular ion CHDBrI$^+$ has been identified as a particularly promising candidate for future experimental searches~\cite{Erez:2022ind,Landau_2023}.\\ 

Finally, measurements of isotope ratios can significantly reduce the  dependence on atomic-structure calculations, which currently constitute a major source of theoretical uncertainty and limit the interpretation of APV experiments~\cite{Ramsey-Musolf:1999qyv,Rosner:1995aj}. 
Such a measurement at the 0.5\% level has been reported in Yb isotopes~\cite{APVISchain}.

%%%%%%%%%%%%%%%%%%%%%%%%%%%%%%%%%%%%%%%%%%%%%%%%%%%%%%%%%%%%%%%%
\subsubsection{Discrete symmetries: CP and T violation}
%%%%%%%%%%%%%%%%%%%%%%%%%%%%%%%%%%%%%%%%%%%%%%%%%%%%%%%%%%%%%%%%

CP violation (CPV), or equivalently T violation (TV) assuming CPT invariance, can be probed through searches for electric dipole moments~(EDMs).
In many cases, these searches are carried out using precision spectroscopy of atoms or molecules subjected to external electric and magnetic fields.
In such setups, CP-violating effects (whether or not they arise directly from an EDM) can be isolated with essentially no SM background, since QED and QCD preserve CP,  and CPV from the weak interaction is far too small to be relevant at the current level of experimental sensitivity. 

CPV may arise from new spin-0 mediators with both CP-even and CP-odd couplings. 
It  may also be induced by new spin-1 mediators through flavor-violating couplings, which are beyond the scope of this review. 
Examples include scalar and pseudoscalar couplings of the form $\phi\bar{\psi}\psi$ and $\phi\bar{\psi}\gamma_5\psi$, as well as gluonic couplings as $\phi GG$ and $\phi G\tilde{G}$. 
At the operator level, CP-violating interactions induce potentials proportional to $\boldsymbol{\sigma} \cdot \hat{\boldsymbol{r}}$. 
The resulting CP-violating effects are conveniently described by the effective potential $V_{PS}$.\\

EDM measurements are typically sensitive to multiple CPV sources. 
For instance, atomic EDMs can probe CP-violating electron-electron and electron-nucleon interactions~\cite{Stadnik:2017hpa}, which may arise from long-range or contact interactions.
Searches for the electron EDM are performed using high-precision spectroscopy of diatomic molecules~\cite{Hudson:2011zz,ACME:2018yjb,Roussy:2022cmp}, exploiting the large internal effective electric fields present in these systems.
Such experiments are sensitive not only to the electron EDM itself, but also to CP-violating electron-electron, electron-nucleon~\cite{Flambaum:2019ejc}, and nucleon-nucleon~\cite{Baruch:2024frj} interactions. 
As a result, they currently provide the most stringent terrestrial constraints on CP-violating new physics for mediator masses $m_\varphi\lesssim10\,\eV$.

%%%%%%%%%%%%%%%%%%%%%%%%%%%%%%%%%%%%%%%%%%%%%%%%%%%%%%%%%%%%%%%%
\subsubsection{Time-dependent signals and dark matter}
%%%%%%%%%%%%%%%%%%%%%%%%%%%%%%%%%%%%%%%%%%%%%%%%%%%%%%%%%%%%%%%%

Atomic clocks can serve as sensitive detectors of ultralight DM~(UDM) with masses far below the eV scale. 
For recent reviews of this approach, see for example Refs.~\cite{Safronova:2017xyt,Antypas:2022asj}. 
In this mass regime, the DM occupation number is very large, so the field is well described as a classical wave exhibiting coherent oscillations at a frequency set by its mass, rather than as a nonrelativistic gas of particles.  
A CP-even scalar field with couplings such as $\phi F_{\mu\nu}F^{\mu\nu}$, $\phi G_{\mu\nu}G^{\mu\nu}$ or $\phi \bar{e}e$, as well as a CP-odd state with quadratic couplings like $\phi^2F_{\mu\nu}F^{\mu\nu}$ can induce time-dependent variations of fundamental constants, see for example Refs.~\cite{Brzeminski:2020uhm,Banerjee:2022sqg,Beadle:2023flm,Kim:2023pvt}. 
These effects manifest themselves either as slow drifts~\cite{Stadnik:2014tta} or as oscillatory modulations at the dark-matter Compton frequency $\omega=m_\phi$~\cite{Arvanitaki:2014faa}. 
Atomic-clock searches are typically sensitive to  variations in the fine-structure constant $\alpha$, the electron-to-proton mass ratio $m_e/m_p$ and the ratio $m_q/\Lambda_{\rm QCD}$.

In these scenarios, the clock signal is linear in the UDM coupling and proportional to local DM energy density, $\rho_{\rm DM}\approx 0.3\,\GeV/\cm^2$~\cite{Read:2014qva}.
This feature distinguishes atomic-clock searches from other probes, which are often quadratic in the couplings and largely independent of the ambient DM density.\\ 

Several spectroscopic strategies have been developed to probe UDM using atomic clocks. 
A primary method involves comparing the frequencies of two clocks with different sensitivities to fundamental constants, using combinations of optical and microwave clocks or pairs of optical clocks. 
Such comparisons enable searches for both oscillatory signals and secular drifts in clock frequency ratios, leading to stringent constraints on UDM-induced variations of fundamental constants~\cite{VanTilburg:2015oza}.

Looking ahead, nuclear clocks based on the transition between the ground state and the low-lying isomeric state of $^{229}$Th offer particularly promising prospects for scalar UDM detection~\cite{Banerjee:2022sqg}, despite theoretical uncertainties stemming from the modelling of the $^{229}$Th nucleus~\cite{Berengut:2009zz,Caputo:2024doz,Beeks:2024xnc}.  
Owing to their enhanced sensitivity to changes in hadronic parameters, nuclear clocks could probe UDM couplings to gluons and quarks more efficiently than conventional atomic clocks. 
Significant recent progress has been made toward the realization of a nuclear clock~\cite{Zhang:2024ngu,Elwell:2024qyh,Tiedau:2024obk}, along with the extraction of the first DM constraints from such systems~\cite{Fuchs:2024xvc}, highlighting their strong potential for future searches.

%%%%%%%%%%%%%%%%%%%%%%%%%%%%%%%%%%%%%%%%%%%%%%%%%%%%%%%%%%%%%%%%
\subsubsection{Data-driven methods: isotope shifts}
\label{sec:DD}
%%%%%%%%%%%%%%%%%%%%%%%%%%%%%%%%%%%%%%%%%%%%%%%%%%%%%%%%%%%%%%%%

Atomic isotope shifts (IS) provide a powerful example of data-driven observables that can be used to probe new physics while largely avoiding theoretical uncertainties. 
An IS refers to the change in atomic transition frequencies arising from differences in neutron number between isotopes, which modify both the nuclear recoil and the nuclear charge distribution. 
These effects make IS particularly sensitive to new SI electron-neutron interactions. 
The IS of transition $i$ between two isotopes with mass numbers $A$ and $A'$ can be expressed to high accuracy as~\cite{King:63,king2013isotope} 
\begin{align}
    \label{eq:IS_theory}
    \nu_i^{A,A'} 
    = 
    K_i \mu^{A,A'} + F_i \drsq^{A,A'} \ .
\end{align}
The first term, known as the \emph{mass shift}, arises from nuclear recoil and is proportional to the change in inverse nuclear mass, $\mu^{A,A'} = (1/m_A - 1/m_{A'})$. 
The second term, the \emph{field shift}, accounts for the penetration of the electronic wave function into the nuclear charge distribution. 
Here, $\drsq^{A,A'}$ denotes the change in the mean-square nuclear charge radius, which typically provides the dominant IS contribution.

Both contributions factorize into transition-independent nuclear quantities, $\mu^{A,A'}$ and $\drsq^{A,A'}$, and isotope-independent electronic coefficients, $K_i$ and $F_i$. 
This separation follows from lowest-order perturbation theory in the electron-nucleus interaction, although the electronic coefficients may incorporate higher-order atomic-structure corrections. 

While IS and nuclear masses can be measured with very high spectroscopic precision, nuclear charge radii remain comparatively less well determined experimentally~\cite{angeli13adndt}. 
However, thanks to the factorized structure above, the dependence on nuclear charge radii can be inferred in a purely data-driven manner by combining IS measurements of multiple transitions. 

By considering two transitions and eliminating $\drsq^{A,A'}$, one obtains a linear relation known as the \emph{King plot}~\cite{King:63},
\begin{align}
    \label{eq:KP}
    \mIS{\nu}_2^{A,A'} 
    = 
    F_{21} \, \mIS{\nu}_1^{A,A'} + K_{21} \ .
\end{align}
Here, $\mIS{\nu}_i^{A,A'} \equiv {\nu_i^{A,A'}}/{\mu^{A,A'}}$ denotes the modified IS, $F_{21} = F_2/F_1$, and $K_{21} = K_2 - F_{21} K_1$. The linearity of this relation has been extensively confirmed experimentally in many atomic systems. 

Crucially, the linearity of the King plot can be exploited as a sensitive probe of new physics~\cite{delaunay17prd,berengut18prl}, see also Refs.~\cite{Frugiuele:2016rii,Berengut:2025nxp}. 
A new SI interaction between neutrons and electrons, mediated by a light boson, induces an additional contribution in Eq.~\eqref{eq:KP} that breaks the linear relation, thereby allowing to constrain its couplings. 
This contribution can be written as $\alphaNP X_{21} \mIS{h}^{A,A'}$, where $\alphaNP = g_{{\rm SI},n}^S g_{{\rm SI},e}^S /( 4\pi)$ for spin-0 mediators and $-g_{{\rm SI},n}^V g_{{\rm SI},e}^V/(4\pi)$ for spin-1 mediators, $X_{21} = X_2 - F_{21} X_1$, and $\mIS{h}^{A,A'} = h^{A,A'} / \mu^{A,A'}$. 
In simple models, $h^{A,A'} = A - A'$ corresponds to the difference in neutron number between the two isotopes, while $X_i$ encodes the effect of the Yukawa potential ${e^{-m_{\phi} r}}/{r}$ on transition $i$ and must be evaluated using atomic many-body theory. 
For sufficiently heavy mediators, the range of the new interaction becomes smaller than the nuclear radius $R_\textrm{nuc}$, rendering the effect indistinguishable from standard nuclear-size contributions~\cite{berengut18prl}. 
As a result, King-plot nonlinearity primarily constrains bosons with masses $m_{\phi} \lesssim  R^{-1}_\textrm{nuc} \sim 100\,\MeV$, while for larger masses the sensitivity rapidly diminishes as $X_{21} \approx 0$.\\

Nonlinearities in King plots have been observed in ytterbium~\cite{Counts:2020aws,Door:2024qqz} and calcium~\cite{Wilzewski:2024wap}, arising from higher-order SM effects such as nuclear deformation, higher-order field shifts, and many-body nuclear correlations. 
These effects limit the sensitivity of IS searches based on a minimal set of transitions and isotopes, as they can mimic the signal of new physics. 
However, this limitation can be overcome in a systematic way. 
One approach is to extend the analysis to additional transitions and isotopes, leading to generalized King relations that allow SM-induced nonlinearities to be isolated and removed~\cite{Berengut:2020itu}. 
Alternatively, the relevant higher-order nuclear effects can be modeled theoretically~\cite{Flambaum:2017onb,Assi:2025nim} and incorporated into the analysis, enabling a separation between Standard Model contributions and genuine new-physics effects. 
Together, these strategies allow IS spectroscopy to retain strong discovery potential despite the presence of observed nonlinearities.

%%%%%%%%%%%%%%%%%%%%%%%%%%%%%%%%%%%%%%%%%%%%%%%%%%%%%%%%%%%%%%%%
\section{Spectroscopic probes of new physics}
\label{sec:systems}
%%%%%%%%%%%%%%%%%%%%%%%%%%%%%%%%%%%%%%%%%%%%%%%%%%%%%%%%%%%%%%%%

\begin{figure}[h]
\includegraphics[width=0.75\textwidth]{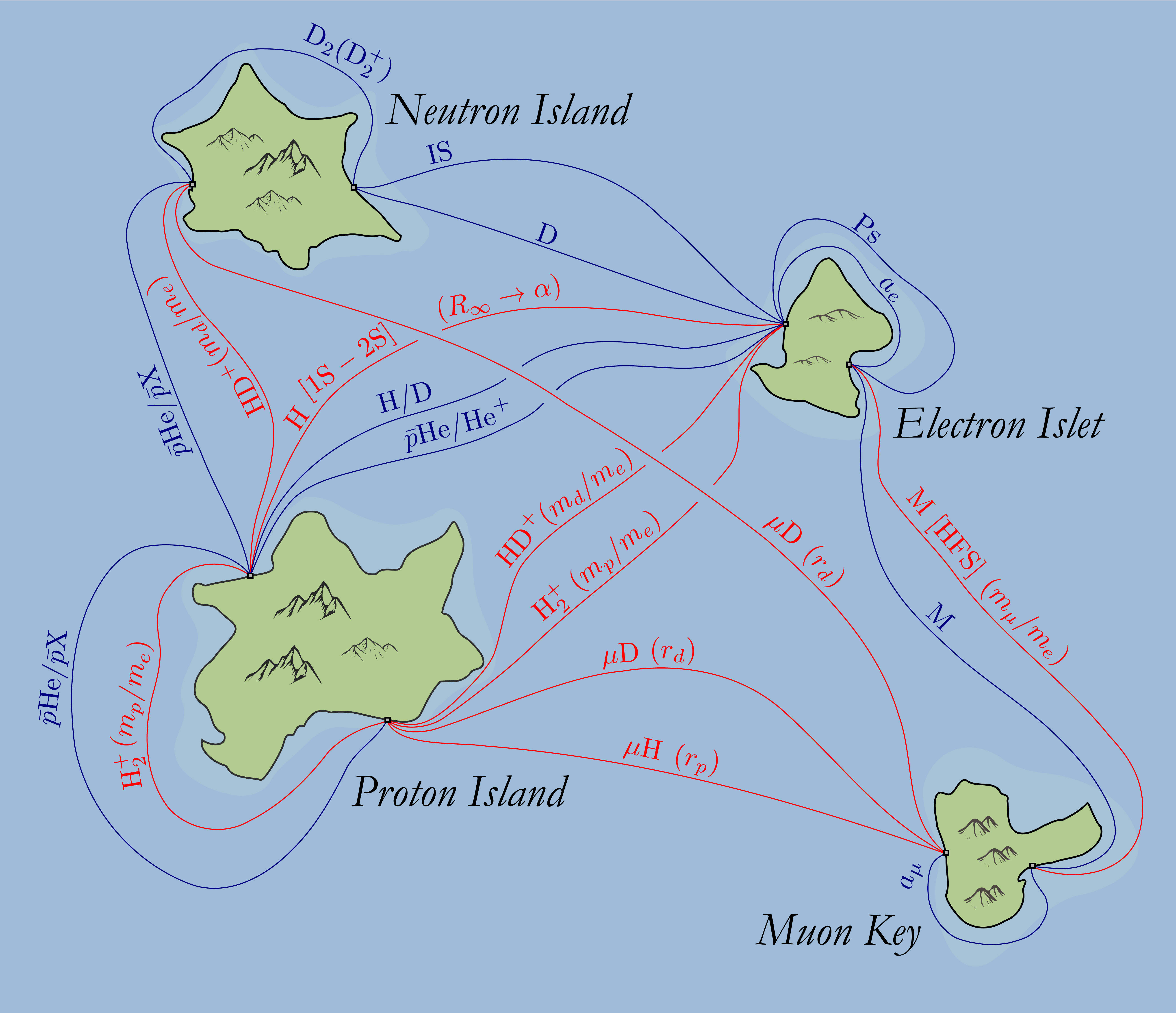}
\caption{Schematic map of effective new-physics couplings to the proton, neutron, electron, and muon, highlighting the spectroscopic systems that connect them. 
Systems containing transitions essential for the determination of fundamental constants are shown in red, with the corresponding constant indicated in parentheses.} 
\label{fig:chart}
\end{figure}

In this section, we review the atomic and molecular systems that have been used as spectroscopic probes of new physics.
We divide these systems into four broad categories: 
few-electron atoms, many-electron atoms, exotic atoms, and molecules. 
For each category, we discuss the relevant energy scales, symmetries and experimental and theoretical techniques, and highlight the most important systems and transitions. 
A summary is provided in Table~\ref{tab:systems}.
As discussed above, BSM physics manifests itself in spectroscopy primarily through shifts in transition frequencies induced by new effective potentials. 
Consequently, spectroscopy experiments are typically sensitive only to products of two new-physics couplings. 
(An important exception is the time variation of fundamental constants, which depends linearly on the new-physics coupling.)
An illustration of this structure is shown in \textbf{Figure~\ref{fig:chart}}, which summarizes the relevant new-physics couplings and indicates which systems are sensitive to particular combinations of them. 

%%%
%%%

%%%
\begin{table}[t]
\centering
\begin{tabular}{ccccc}
\hline\hline
System & Transition & $u_r^{\rm exp}$ & $u_r^{\rm th}$ & Relative range  \\
\hline
H & $1S-2S$ & $4.2 \times 10^{-15}$~\cite{Parthey:2011lfa} & $5.7\times 10^{-13}$~\cite{Mohr:2024kco} &  \\
H & $1S-3S$ & $2.3 \times 10^{-13}$~\cite{Grinin:2020txk} & $5.7\times 10^{-13}$~\cite{Mohr:2024kco} &  \\
H & $2S-6P$ & $6.6 \times 10^{-13}$~\cite{Maisenbacher:2026nau} & $3.1\times 10^{-13}$~\cite{Maisenbacher:2026nau} &  \\
H/D IS& $1S-2S$ & $2.2 \times 10^{-11}$~\cite{Parthey:2010aya} & $5.1 \times 10^{-10}$~\cite{Mohr:2024kco} &  \\
$\mu$H & $2S-2P$ & $1.1\times 10^{-5}$~\cite{Antognini:2013txn} & $1.2\times 10^{-5}$~\cite{Pachucki:2022tgl} & $m_e/m_\mu$ \\
$\mu$D & $2S-2P$ & $1.7\times 10^{-5}$~\cite{CREMA:2016idx} & $1.0\times 10^{-4}$~\cite{Pachucki:2022tgl} & $m_e/m_\mu$ \\
Ps & $1S-2S$ & $2.6 \times 10^{-9}$~\cite{Fee:1993zz} & $8\times10^{-10}$~\cite{Cassidy:2018tgq} & $2$ \\
M & $1S-2S$ & $4.0 \times 10^{-9}$~\cite{Meyer:1999cx} & $5.7 \times 10^{-10}$~\cite{Jungmann:2001zz} & \\
$\bar{p}$He & Rydberg & $3 \times 10^{-9}$~\cite{ASACUSA:2011lsy,ASACUSA:2016xeq} & $4.7 \times 10^{-11}$~\cite{Korobov:2015yya} & $\sim 1/2$ \\
H$_2^+$ & rovib. & $8.0 \times 10^{-12}$~\cite{Alighanbari:2025} & $8.0 \times 10^{-12}$~\cite{Korobov:2021} & $2$ \\
HD$^+$ & rovib. & $3 \times 10^{-12}$~\cite{Patra:2020brw,Alighanbari:2023} & $8.0 \times 10^{-12}$~\cite{Korobov:2021} & $2$ \\
$^4$He & $2\,^3S_1-2\,^3P$ & $3.1 \times 10^{-12}$~\cite{Wen:2024xbe} & $2.0 \times 10^{-10}$~\cite{Patkos:2021wam} &  \\
$^4$He & $2\,^3P_0-3\,^3D_1$ & $5.5 \times 10^{-11}$~\cite{Luo:2016} & $3.1 \times 10^{-12}$~\cite{Patkos:2021wam} &  \\
D$_2$ & $D_0$& $7.1 \times 10^{-10}$~\cite{Hussels:2022} & $7.1 \times 10^{-10}$~\cite{Puchalski:2019} & $1.4$ \\
 $\bar{p}^{20}$Ne & $6H-5G$ & $\sim10^{-4}$~\cite{Gotta:1999vj} & $<10^{-4}$~\cite{Paul:2020cnx} & $0.0003-0.03$ \\
\hline\hline
H & $\Delta_{12}$ & $1.4 \times 10^{-4}$~\cite{Bullis:2023big} & $4.7 \times 10^{-5}$~\cite{Yerokhin:2008} &  \\
D & $\Delta_{12}$ & $5.0 \times 10^{-3}$~\cite{Kolachevsky:2004} & $4.4 \times 10^{-5}$~\cite{Karshenboim:2001ia} &  \\
$^3$He$^+$ & $\Delta_{12}$ & $6.0 \times 10^{-5}$~\cite{Prior:1977zz,Schneider:2022mze} & $4.2 \times 10^{-5}$~\cite{Yerokhin:2008} & $1/2$ \\
Ps & $1\,^3S_1-1\,^1S_0$ & $3.6 \times 10^{-6}$~\cite{Ritter:1984mqy} & $2.3 \times 10^{-6}$~\cite{Adkins:2022omi} & $2$ \\
$^4$He & $2\,^3P_1-2\,^3P_2$ & $1.1 \times 10^{-8}$~\cite{Kato:2018uhr} & $7.4 \times 10^{-7}$~\cite{Pachucki:2010zz} &  \\
H$_2$ & HFS & $4.2 \times 10^{-4}$~\cite{Harrick:1953} & $\ll u_r^{\rm exp}$~\cite{Code:1971} & $1.4$ \\
HD & HFS & $2.3 \times 10^{-5}$~\cite{Ledbetter:2012xd} & $1.8 \times 10^{-3}$~\cite{Puchalski:2018} & $1.4$ \\
\hline\hline
\\
\end{tabular}
\caption{Summary of simple atomic, molecular, and exotic systems used in spectroscopy-based searches for new physics.
The top section lists systems best probing spin-independent interactions, while the bottom section lists those most sensitive to spin-dependent interactions.
The quantities $u_r^{\rm exp}$ and $u_r^{\rm th}$ denote the relative experimental and theoretical uncertainties, respectively. The last column indicates the shortest interaction range probed, expressed in units of the Bohr radius $a_0\sim(\alpha m_e)^{-1}$ whenever it is different from $a_0$. 
For molecular systems, the indicated ranges correspond to the equilibrium bond length and are relevant for nucleon-nucleon interactions.}
\label{tab:systems}
\end{table}
%%%

%%%%%%%%%%%%%%%%%%%%%%%%%%%%%%%%%%%%%%%%%%%%%%%%%%%%%%%%%%%%%%%%
\subsection{Atomic systems with one or two electrons} 
\label{sec:fewelectrons}
%%%%%%%%%%%%%%%%%%%%%%%%%%%%%%%%%%%%%%%%%%%%%%%%%%%%%%%%%%%%%%%%

As discussed in Section~\ref{sec:th-exp}, the primary advantage of few-electron atomic systems is that their electronic structure can be calculated with exceptionally high precision, in some cases matching or exceeding current experimental accuracy. 
The state-of-the-art theoretical and experimental precision for the relevant systems is summarized in Table~\ref{tab:systems}. \\

The most precise comparisons between theory and experiment have been achieved in  hydrogen and deuterium spectroscopy. 
For these systems, theoretical predictions have reached sub-parts-per-trillion accuracy and are currently limited by uncalculated higher-order QED effects, including two-loop~\cite{Yerokhin:2024cyw}, three-loop~\cite{Karshenboim:2019siz}, and radiative-recoil~\cite{Mohr:2024kco,Yerokhin:2018gna} contributions. 
Experimentally, the very narrow $1S-2S$ transition, whose natural linewidth is $\sim1.3\,$Hz, has enabled measurements at the few-$10^{-15}$ level~\cite{Parthey:2011lfa}. 
Other transitions possess much larger natural widths, typically in the MHz range, which fundamentally limits the achievable precision. 
Nevertheless, through careful control of systematic effects, experimental uncertainties at the level of $\sim 10^{-4}$ of the linewidth have been attained~\cite{Beyer:2017gug}.

Energy levels in hydrogen and deuterium are primarily sensitive to the Rydberg constant $R_\infty$ and to the proton and deuteron charge radii $r_p$ and $r_d$, respectively. 
Sensitivities to the fine-structure constant and to the electron-proton mass ratio are comparatively small and can be taken from independent measurements. 
Given that nuclear charge radii are now determined with high precision from muonic atom spectroscopy, assuming the absence of BSM muon interactions, the $1S-2S$ transition of hydrogen effectively fixes  the Rydberg constant. 
Other measured transitions, such as the $1S-3S$ and $2S-6P$ transitions listed in Table~\ref{tab:systems}, then provide constraints on new SI electron-proton and electron-neutron interactions. 

Measurements of the same transition, in particular $1S-2S$~\cite{Parthey:2010aya}, in both hydrogen and deuterium provide access to isotope shifts, which benefit from reduced theoretical uncertainties due to cancellation of common contributions~\cite{Jentschura:2011mdl}. 
As discussed in Section~\ref{sec:th-exp}, isotope shifts are particularly sensitive to new electron-neutron interactions.\\

Spectroscopy of helium and helium-like systems offers complementary probes. 
Precision measurements of helium transition frequencies is the most sensitive probe of electron-electron interactions, in particular since theoretical predictions for triplet S and P states were recently improved through a complete calculation of QED corrections of order $m_e\alpha^7$~\cite{Patkos:2021wam}. 
However, the theoretical frequencies of the $2^3S_1-3^3D_1$ and $2^3P_0-3 ^3D_1$ transitions disagree  with experimental measurements by more than $10\,\sigma$~\cite{Dorrer:1997,Luo:2016}, whereas there is good agreement  between theory and experiment for the $2^3S-2^3P$ transition~\cite{Wen:2024xbe}. 
This discrepancy has been confirmed by measurements of the ionization threshold of the $2^3S_1$ level~\cite{Clausen:2025}. 
It is likely caused by a yet unidentified problem in the theoretical calculation of the $m_e\alpha^7$ QED corrections that would induce a common shift to the energies of the $2^3S$ and $2^3P$ levels. 
Once this discrepancy is resolved, helium spectroscopy will provide the strongest constraint on SI electron-electron interactions for mediator masses below $\sim 1\,\keV$.\\

Spin-dependent BSM interactions can be probed through measurements of hyperfine structure intervals, see for example Ref.~\cite{Cong:2024qly} for a review. 
In hydrogen and deuterium, the hyperfine splitting of the $1S$ and $2S$ states have been measured with high precision; 
 see Ref.~\cite{Karshenboim:2005iy}, and Ref.~\cite{Bullis:2023big} for a recent improvement of the $2S$ measurement. 
Although theoretical predictions for individual hyperfine splittings are significantly limited by nuclear-structure uncertainties,  as discussed in Section~\ref{sec:th-exp} most nuclear contributions cancel in the specific difference
\begin{align}
    \Delta_{12} \equiv \Delta E_{\rm hfs} (1S) - 8 \Delta E_{\rm hfs} (2S)\,,
\end{align} 
which can be calculated with significantly improved accuracy~\cite{Karshenboim:2005iy}. 
This observable currently provides the strongest atomic-spectroscopy constraints on new SD electron-proton  interactions~\cite{Karshenboim:2010cg,Karshenboim:2010cm,Cong:2024nat} arising in the $V_{AA}$, $V_{VV}$, and $V_{PP}$ potentials. 
Similar measurements in deuterium and $^3$He$^+$ extend these constraints to electron-neutron interactions. In addition, the fine structure of the $2^3P$ state in helium constrains SD electron-electron interactions.\\

Looking ahead, further theoretical improvements in hydrogen and helium atoms remain possible, though increasingly challenging~\cite{Yerokhin:2024cyw,Karshenboim:2019siz}. 
On the experimental side, substantial gains are limited by natural linewidths, with the notable exception of the $1S–2S$ transition, where proposals toward an optical clock may enable further progress~\cite{Hydrogenclock}.
New experimental efforts on systems such as helium ions~\cite{Moreno:2023amv,Grundeman:2023juv}, as well as the resolution of existing discrepancies in helium spectroscopy~\cite{Patkos:2021wam}, offer promising directions for future advances.

%%%%%%%%%%%%%%%%%%%%%%%%%%%%%%%%%%%%%%%%%%%%%%%%%%%%%%%%%%%%%%%%
\subsection{Atomic systems with many electrons} 
\label{sec:HeavyAtoms}
%%%%%%%%%%%%%%%%%%%%%%%%%%%%%%%%%%%%%%%%%%%%%%%%%%%%%%%%%%%%%%%%

Heavy atoms with many electrons offer two important advantages compared to light few-electron systems.
First, the vast range of available elements and electronic configurations across the periodic table provides access to a much wider set of systems and transitions. 
This flexibility (rather than the complexity of the electronic structure itself) enables the selection of exceptionally narrow and environmentally robust transitions, allowing experimental accuracies to reach the $\sim 10^{-19}$ level in state-of-the-art atomic clocks. 
Second, many SM and BSM effects are strongly enhanced in heavy atoms, in some cases scaling as  $\sim Z^3$, which significantly increases their sensitivity to new interactions.\\

These advantages come at the cost of reduced theoretical control, since \textit{ab initio} SM predictions for many-electron systems remain substantially less precise than for light atoms. 
As a result, BSM searches in heavy atoms typically rely on observables that minimize or eliminate theoretical input. 
As discussed in Section~\ref{sec:TheoryFree}, this includes symmetry-violating observables, isotope-shift measurements analyzed through King plots, and time-dependent signals.\\

Atomic parity violation has been measured most precisely in $^{133}$Cs~\cite{Wood:1997zq} (see also~\cite{Guena:2004sq}), reaching a relative accuracy of about $0.4\%$, with complementary measurements in $^{205}$Tl~\cite{Edwards:1995zz,Vetter:1995vf}, $^{208}$Pb~\cite{Meekhof:1993zz} and $^{209}$Bi~\cite{Macpherson:1991opp}. 
These experiments probe parity-violating electron-nucleon interactions, parameterized by the axial-vector potential $V_{AV}$, and constrain the product $g_e^Ag_N^V$. 
Using the caesium measurement, one obtains~\cite{Bouchiat:2004sp,Dzuba:2017puc}
\begin{align}
    \label{eq:APVBound}
    g^A_e\left(g^V_d+\frac{188}{399}g^V_u\right)
    \lesssim 
    \frac{0.5}{K(m_X)} \left(\frac{m_X^2}{\TeV}\right)^2
\end{align}
where $K(m_X\lesssim 0.1\,\MeV)\approx 0.025$ and $K(m_X\gtrsim 0.1\,\GeV)\approx 0.98$~\cite{Bouchiat:1983uf,Bouchiat:2004sp}, which is the strongest bound from atomic systems. 
The dependence on a specific linear combination of quark couplings implies the existence of blind directions, which can be lifted by APV measurements in different atomic species, making even lower-precision measurements highly complementary.

Heavy atoms also offers exceptional sensitivity to CP-violating electron-nucleon and electron-electron interactions through searches for permanent EDMs. 
EDM measurements in systems such as Cs, Tl, Yb$^+$, Hf$^{3+}$ and Th$^+$ currently set the most stringent terrestrial bounds on a wide class of CP-violating operators~\cite{Stadnik:2017hpa}; see also Ref.~\cite{Flambaum:2019ejc} for a review of short-ranged hadronic CPV.

IS spectroscopy and King-plot analyses in heavy atoms provide a data-driven method to search for new SI electron-neutron interactions, largely avoiding atomic-structure uncertainties (see Section~\ref{sec:DD}).
High-precision IS measurements have been performed in Yb and Yb$^+$ at the $5\,$Hz ($\sim 4\times 10^{-9}$) level~\cite{Door:2024qqz}, and in Ca, Ca$^+$ and Ca$^{15+}$ at sub-Hz ($\sim 10^{-10}$) level~\cite{Wilzewski:2024wap}, with recent experiments reaching the 10\,mHz using entangled isotopes in Sr$^+$~\cite{Manovitz:2019czu} and multiple isotope pairs in Ca$^+$~\cite{Wilzewski:2024wap}.
Significant King-plot nonlinearities have been observed in Yb and Ca systems, which are however currently attributed to higher-order nuclear and field-shift effects expected within the SM rather than to new physics. 
Generalized King analysis~\cite{Fuchs:2025lyr} and the ZigZag~\cite{Counts:2020aws} methods likewise show no preference for a new boson, in agreement with bounds derived from helium-deuterium IS and complementary constraints from neutron scattering and $(g-2)_e$ data~\cite{berengut18prl,delaunay17prd,Frugiuele:2016rii}. 
A first step toward a global analysis of IS searches for new physics has recently been presented in Ref.~\cite{Fuchs:2025lyr}.\\

Finally, there are several proposals to perform high-precision spectroscopy of Rydberg states in many-electron atoms~\cite{Ramos:2017}. 
Owing to their large principal quantum numbers, these energy levels can be calculated with high accuracy using quantum-defect theory. 
While the primary goal of such measurements is the determination of the Rydberg constant, achieving sufficient experimental precision would also make these systems sensitive to potential new-physics contributions to the energy levels~\cite{Jones:2019qny}. 

%%%%%%%%%%%%%%%%%%%%%%%%%%%%%%%%%%%%%%%%%%%%%%%%%%%%%%%%%%%%%%%%
\subsection{Exotic atoms}
%%%%%%%%%%%%%%%%%%%%%%%%%%%%%%%%%%%%%%%%%%%%%%%%%%%%%%%%%%%%%%%%

Exotic atoms provide unique laboratories for probing physics beyond the SM by replacing the bound electron with a heavier particle or by forming purely leptonic bound states. 
We divide these systems into three broad classes.
First, \emph{purely leptonic atoms}, such as positronium and  muonium, which are bound states of  $e^+e^-$ and $\mu^+e^-$, respectively. 
In these systems, theoretical predictions are exceptionally clean, being governed almost entirely by QED and free from nuclear-structure effects. 
Second, \emph{muonic atoms}, in which a muon replaces an electron, as in muonic hydrogen or deuterium. 
While the underlying theory remains under good control, nuclear effects are strongly enhanced, particularly for $S$ states, due to the much smaller Bohr radius. 
Third, \emph{hadronic atoms}, where an antiproton, negatively charged pion or kaon replaces the electron, as in $\bar{p}$He and $\bar{p}$Pb.\\ 

Muonic and hadronic atoms differ from ordinary electronic atoms in two important ways. 
First, the orbiting particle is much heavier than the electron, reducing the effective Bohr radius by a factor of $\cO(m_e/m_{\mu,p})$. 
As a result, these systems probe much shorter distance scales and are sensitive to new physics with mediator masses in the MeV range. 
Second, they provide access to  new interactions involving muons and hadrons, or purely hadronic couplings, which are largely invisible in conventional atomic spectroscopy.
Theoretical challenges vary across the different classes. 
Muonic and hadronic systems require precise knowledge of nuclear parameters such as charge radii, which are often extracted from the same measurements~\cite{Antognini:2013txn,CREMA:2016idx,Krauth:2021foz,CREMA:2025zpo,Beyer:2025imi}.  
All exotic atoms suffer from short lifetimes and low statistics that limit experimental uncertainties.

\subsubsection{Purely leptonic atoms}
High-precision spectroscopy has been achieved in two purely leptonic systems: positronium and muonium. 
A comprehensive analysis of their sensitivity to new physics was presented in Ref.~\cite{Frugiuele:2019drl}; see also Table~\ref{tab:systems}.
Positronium is sensitive exclusively to electronic couplings, while muonium probes the product of electronic and muonic couplings of a new mediator. 
In both systems, the $1S-2S$ transition has been measured at the $\sim10^{-9}$ level, with theoretical predictions known to even higher precision. 
The resulting constraint from positronium $1S-2S$ on purely electronic couplings is only a factor of a few weaker than that derived from the $(g-2)_e$ in the $m_\varphi\to0$ limit~\cite{Delaunay:2017dku}.
For muonium, the $1S-2S$ transition yields bounds on $g_{{\rm SI},e} g_{{\rm SI},\mu}$ that are stronger than those obtained by naively combining $(g-2)_e$ and $(g-2)_\mu$ constraints in the same limit. 
There are ongoing efforts to improve the measurements in both positronium and muonium~\cite{Borges:2025qfk,Javary:2024gca} and develop the theory further~\cite{Adkins:2022omi}.
Hyperfine structure~(HFS) intervals have also been measured in both systems and can be used to constrain SD interactions, although these bounds are typically weaker than those from magnetic-moment measurements. 
In the case of muonium, the sensitivity is additionally limited by the precision with which the muon-to-electron mass ratio is known. 
Alternatively, the $m_e/m_\mu$ can be extracted from the $1S-2S$ transition in muonium, and when combined with the hyperfine splitting, it enables an independent, precise determination of $(g-2)_\mu$ based purely on spectroscopic data~\cite{Delaunay:2021uph}. 

\subsubsection{Muonic atoms}
In muonic atoms, the characteristic length scale is reduced to $a_0(m_e/m_\mu)\sim (0.8\,\MeV)^{-1}$, enabling sensitivity to significantly heavier mediators than in electronic atoms. 
Precision measurements of transitions in muonic hydrogen and muonic deuterium, notably the $1S-2S$ transitions measured at the $\sim10^{-5}$ level, have been used both to determine the proton and deuteron charge radii and to constrain muon-nucleon interactions, see for example Ref.~\cite{Frugiuele:2021bic}. 
Disentangling these effects requires a global analysis combining electronic and muonic spectroscopy, as discussed in Ref.~\cite{Delaunay:2022grr}. 
Additional constraints arise from transitions in heavier muonic atoms. 
For instance, the $3D_{5/2}-3P_{3/2}$ in $^{24}$Mg and $^{28}$Si~\cite{Beltrami:1985dc} probe muon-nucleon couplings with peak sensitivity around $m_\varphi\sim 1\,\MeV$, while measurements of the $1S$ hyperfine splitting in $\mu{}^7$Li at the percent level constrain  axial-vector interactions~\cite{Ruckstuhl:1985xg}. 
For mediator masses $m_X\gtrsim 3\,\MeV$, the resulting bound on SD couplings is $g^V_{{\rm SD},\mu}g^V_{{\rm SD},N}<10^{-2}(m_X/\GeV)^2$, tightening to $g^V_{{\rm SD},\mu}g^V_{{\rm SD},N}<10^{-6}$ in the massless limit. 
Moreover, $2S-2P$ transitions in $\mu^4$He and $\mu^3$He have been measured with few-$10^{-5}$ relative accuracy~\cite{Krauth:2021foz,CREMA:2025zpo}. 
While these measurements primarily determine the $\alpha$-particle and helion charge radii, in combination with electronic helium spectroscopy or scattering data they can also constrain muon-nucleon interactions.

\subsubsection{Hadronic atoms}
Hadronic atoms are generally challenging for BSM searches due to poorly controlled strong-interaction effects. 
However, circular states with large principal quantum number $n$ and maximal orbital angular momentum $l=n-1$ provide a notable exception. 
In these states, the overlap between the hadron and the nucleus is strongly suppressed, significantly reducing hadronic uncertainties, while the small Bohr radius still allows sensitivity to MeV-scale new physics. 
Transition energies in such systems can be predicted at the $10^{-6}$ level or better, up to nuclear polarization effects~\cite{Paul:2020cnx,Baptista:2025kai,Patkos:2025vxd}. 
Reference~\cite{Liu:2025ows} demonstrated how measurements of two transitions can be used both to constrain new SI hadronic interactions and to extract nuclear polarization effects directly from data. 
The PAX experiment~\cite{Baptista:2025kai} aims to measure circular transitions in antiprotonic atoms, including Xe and other elements, with relative precision at the $10^{-6}$ level.

%%%%%%%%%%%%%%%%%%%%%%%%%%%%%%%%%%%%%%%%%%%%%%%%%%%%%%%%%%%%%%%%
\subsection{Molecules}
%%%%%%%%%%%%%%%%%%%%%%%%%%%%%%%%%%%%%%%%%%%%%%%%%%%%%%%%%%%%%%%%

Molecular hydrogen ions, like HD$^+$ and H$_2^+$, provide a complementary class of precision systems for probing physics beyond the SM, with sensitivities that are often different from those of atomic systems. 
Compared to atoms, molecular spectra involve additional rotational and vibrational degrees of freedom linked to the relative motion of the nuclei, which introduce both experimental advantages and theoretical challenges. 

From a theoretical standpoint, HD$^+$ and H$_2^+$ ions are the simplest molecular systems and represent the natural extension of the hydrogen atom. 
Their energy levels can be calculated with high precision~\cite{Korobov:2021}, although the theoretical uncertainty is typically about an order of magnitude larger than in atomic hydrogen. 
As a result, these systems are not competitive for constraining electron–proton or electron–neutron interactions. 
Their primary strength instead lies in their sensitivity to nucleon–nucleon interactions, which are difficult to access in atomic systems~\cite{Salumbides:2013aga}. 

Experimentally, molecular hydrogen ions are particularly attractive because their rovibrational transitions have extremely narrow natural linewidths. 
These ions can be trapped and sympathetically cooled to temperatures of order $10\,$mK using laser-cooled atomic ions, enabling spectroscopy in the Lamb–Dicke regime and suppressing first-order Doppler broadening. 
To date, four rovibrational transitions in HD$^+$~\cite{Alighanbari:2020,Patra:2020brw,Kortunov:2021,Alighanbari:2023} and one in H$_2^+$~\cite{Alighanbari:2025} have been measured with relative uncertainties in the $10^{-11}-10^{-12}$ range. 
These measurements provide some of the strongest spectroscopic constraints on SI proton–proton and proton–neutron interactions. 
Similar measurements in the D$_2^+$ molecular ion would enable comparable sensitivity to neutron–neutron interactions. 
At present, the best spectroscopic bounds in that sector come from the dissociation energy of neutral D$_2$~\cite{Salumbides:2013aga,Hussels:2022,Puchalski:2019}. 
In the future, the precision in molecular hydrogen ions may be significantly improved further using quantum-logic spectroscopy techniques~\cite{Holzapfel:2024rmv} and measuring well-chosen ensembles of transitions in several isotopologues~\cite{Karr:2025}. \\

SD nucleon–nucleon interactions are most stringently constrained by hyperfine structure measurements in neutral hydrogen molecules. 
In H$_2$, proton–proton interactions have been bounded by comparing experimental data~\cite{Harrick:1953} with theoretical predictions~\cite{Code:1971} for the tensor (dipole–dipole) spin–spin coupling coefficient~\cite{Ramsey:1979bzw}. 
In the HD molecule, high-precision NMR measurements have resolved the extremely small scalar proton–deuteron spin–spin coupling, which arises as a second-order effect mediated by the electron. 
Comparison with theory yields constraints on SD electron–neutron interactions~\cite{Ledbetter:2012xd,Puchalski:2018}, which can be interpreted in terms of axial–vector potentials of the form given in Eq.~\eqref{eq:VAV}.

PV nucleon–nucleon interactions, which are also of SD type, can be probed in chiral molecules by comparing the hyperfine structure of opposite enantiomers~\cite{Baruch:2024fbh}. 
In such systems, vibrational spectra are sensitive to Standard Model PV effects and to PV electron–nucleon interactions from new physics, while hyperfine transitions are expected to be largely free of SM backgrounds. 
In particular, sub-Hz precision spectroscopy of the hyperfine structure in CHDBrI$^+$ has been proposed as a leading probe of new spin-1 mediators with both axial and vector couplings to neutrons.
The vibrational modes probe the SM PV effects and new physics PV electron-nucleon interactions, with a reach comparable to APV.

Finally, diatomic molecules provide unparalleled sensitivity to CP-violating interactions through measurements of the electron EDM. 
Precision spectroscopy in heavy polar molecules~\cite{Hudson:2011zz,ACME:2018yjb,Roussy:2022cmp} is sensitive not only to the electron EDM but also to CP-violating electron–nucleon, electron–electron~\cite{Stadnik:2017hpa}, and nucleon–nucleon~\cite{Baruch:2024frj} interactions, described by the effective potential $V_{PS}$. 
These experiments currently yield the most stringent terrestrial bounds on CP-violating new physics for mediator masses in the range $1\, \eV\lesssim m_\varphi \lesssim 10\,\keV$. 

%%%
\begin{figure}[t]
\includegraphics[width=0.8\textwidth]{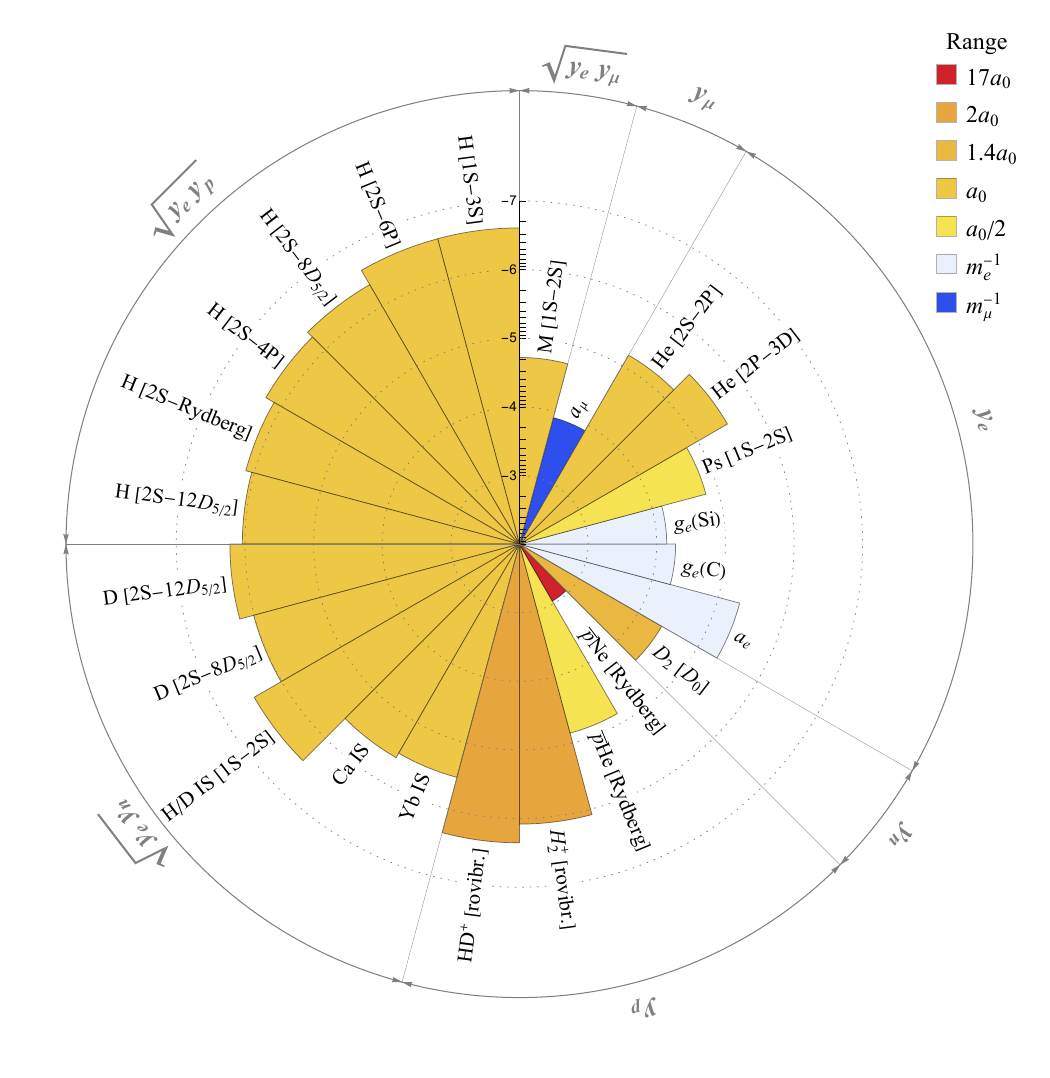}
\caption{Sensitivity reach of precision spectroscopic systems to new physics.
The shaded sectors illustrate the sensitivity of various transitions in atomic, molecular, and exotic systems and to different combinations of effective new-physics couplings to electrons, muons, protons, and neutrons. The sectors are organized based on specific (products of) couplings probed by each systems, highlighting the complementarity between hydrogenic atoms, molecular ions, helium, positronium, muonium, and isotope-shift measurements. The radial axis encodes the strength of the probed coupling combinations, while the color shading indicates the characteristic interaction range.}
\label{sensitivities}
\end{figure}
%%%

%%%%%%%%%%%%%%%%%%%%%%%%%%%%%%%%%%%%%%%%%%%%%%%%%%%%%%%%%%%%%%%%
\section{Combined analysis of spectroscopy data}
\label{sec:codata-np}
%%%%%%%%%%%%%%%%%%%%%%%%%%%%%%%%%%%%%%%%%%%%%%%%%%%%%%%%%%%%%%%%

Here, we perform a global analysis of precision spectroscopy data following the methodology of Ref.~\cite{Delaunay:2022grr}.
We use the CODATA-2022~\cite{Mohr:2024kco} dataset together with the state-of-the-art theoretical predictions and include additional experimental results on electronic hydrogen~\cite{Maisenbacher:2026nau}, muonium~\cite{Meyer:1999cx}, electronic~\cite{Dorrer:1997,Pastor:2012iib,Luo:2016,Rengelink:2018zyd,Wen:2024xbe,vanderWerf:2025ovr}, muonic~\cite{CREMA:2025zpo} and antiprotonic~\cite{ASACUSA:2011lsy,ASACUSA:2016xeq} helium, HD$^+$~\cite{Alighanbari:2023} and H$_2^+$~\cite{Alighanbari:2025} and related theoretical predictions~\cite{Jungmann:2001zz,Patkos:2021wam,Pachucki:2025,Pachucki:2022tgl,Korobov:2015yya}. 
Our dataset is dominated by few-electron atoms (see Section~\ref{sec:fewelectrons}) and other simple systems, for which theoretical calculations are under excellent control (see Section~\ref{sec:th-exp}). 
These systems form the backbone of the CODATA determination of fundamental constants and are ideally suited for a simultaneous extraction of constants and constraints on new physics. 
The framework of Ref.~\cite{Delaunay:2022grr} allows for a consistent treatment of potential new interactions alongside the determination of fundamental constants, avoiding biases that could arise from fixing constants to their SM values.\\

We present results for the following benchmark models:
\begin{itemize}
\item \emph{Dark photon}~\cite{Holdom:1985ag,Okun:1982xi}: 
    a spin-1 state with vector couplings to the SM fermions proportional to their electric charges $q_\psi$, namely $g^V_\psi=\varepsilon e q_\psi$, where $e\equiv \sqrt{4\pi\alpha}$, and $g^A_\psi=0$. 
    Such a state typically originates from an additional broken U(1)$_X$ gauge symmetry, with interactions with the SM fields mediated by the renormalizable kinetic-mixing term $-\frac{\varepsilon}{2}X_{\mu\nu}F^{\mu\nu}$.
\item \emph{$B-L$ gauge boson}~\cite{Davidson:1978pm,Marshak:1979fm}: 
    a spin-1 state coupled to baryon minus lepton number, with purely vector couplings, $g^V_{p,n}=-g^V_{e,\mu}\equiv g_{B-L}$ and $g^A_\psi=0$.
\item \emph{Higgs portal scalar}~\cite{Patt:2006fw,OConnell:2006rsp}: 
    a light CP-even spin-0 mixing with the SM Higgs, leading to Yukawa-like couplings to fermions, $\kappa_\psi=(f_\phi/v)\sin\theta$, where $\theta$ is the mixing angle and $v\approx246\,\GeV$ is the vacuum expectation value of the SM Higgs, and $\tilde{\kappa}_\psi=0$. The effective coupling to nucleons is dominated by the $\phi GG$ term, yielding
 $g_N^S\approx 0.31\sin\theta (f_\phi/v)$ and $g_N^P=0$.
\item \emph{Featheron scalar}~\cite{Delaunay:2022grr,Delaunay:2025lhl}: 
    a CP-even spin-0 state with Yukawa-like couplings exclusively to fermions lighter than $\sim 1\,\GeV$, namely the $q=u,d,s$ quarks and $\ell=e,\mu$ with $\kappa_{q}=\kappa_ \ell=1$, and all other couplings vanishing. 
    This structure of couplings evades stringent constraints from $K\to\pi\phi$ decays  while allowing observable signature in precision spectroscopic measurements.
Furthermore, the featheron scalar is assumed to decay predominantly  into dark sector states, thereby avoiding bounds from beam-dump experiments. 
\end{itemize}
In all cases, the leading effect on spectroscopy arises from SI interactions described by the $V_{SS}$ and $V_{VV}$ potentials.
For convenience, we present constraints on the effective coupling combination $\alpha_\varphi\equiv g_{{\rm SI},e} g_{{\rm SI},p}/(4\pi)$, which is typically probed by atomic systems.\\

\begin{figure}[t]
    \centering
    \begin{subfigure}[b]{0.49\textwidth}
        \includegraphics[width=1\textwidth]{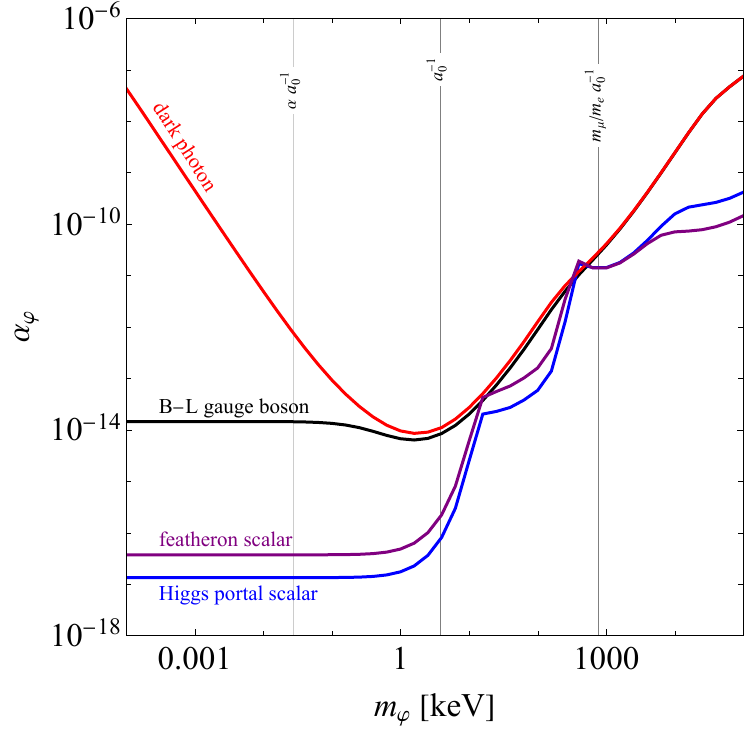}
    \end{subfigure}
    \begin{subfigure}[b]{0.49\textwidth}
        \includegraphics[width=1\textwidth]{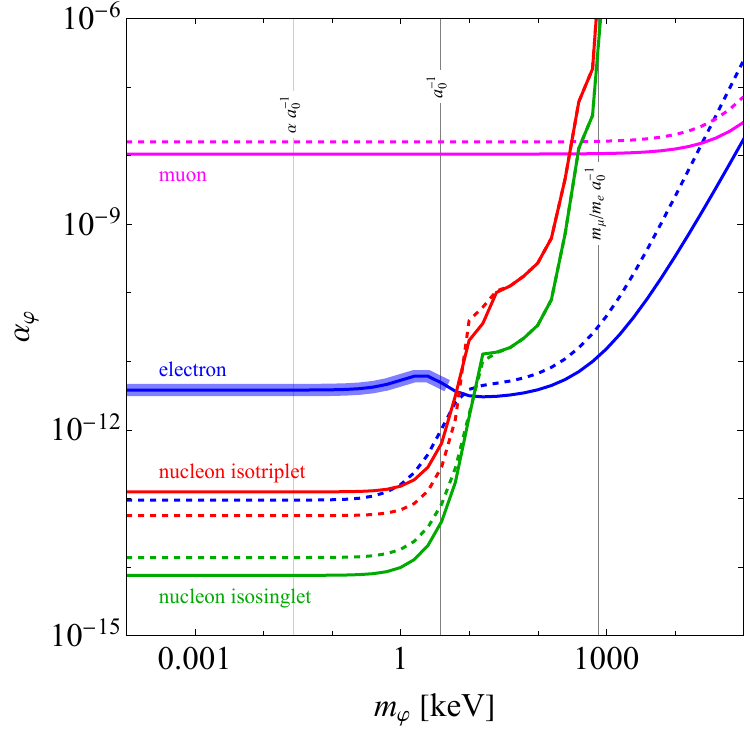}
    \end{subfigure}
    \caption{
    Bounds on spin-independent interactions obtained from a global fit to precision spectroscopic data, simultaneously wih a self-consistent determination of the relevant fundamental constants (see Section~\ref{sec:codata-np}). The left panel shows the 99$\%$ CL constraints on $\alpha_{\varphi}\equiv g_{{\rm SI},e}g_{{\rm SI},p}/(4\pi)$ for four benchmark models, namely a dark photon (red), a $B-L$ gauge boson (black), a Higgs portal  scalar (blue) and a featheron scalar (purple). 
    The right panel shows the 99$\%$ CL constraints on individual spin-independent couplings to the electron (blue), muon (magenta) and nucleon assuming either isotriplet (red) or isosinglet (green) structure. Solid (dashed) curves correspond to a spin-0 mediator $\phi$ (spin-1 mediator $X_\mu$). 
    Thicker bands indicate the mediator-mass ranges for which the fit prefers new physics over the SM at more than $4\sigma$.}
    \label{fig:models}
\end{figure}

The resulting 99\%\,CL bounds shown in Fig.~\ref{fig:models} (left panel). 
For the dark photon and $B-L$ models, the data are consistent with the absence of new physics across the full mediator mass range. 
In contrast, for the Higgs portal and featheron models, a statistically significant excess $\sim 2.6\sigma$ appears for mediator masses around the MeV scale. 
This excess is primarily driven by the $2S-8D$~\cite{Brandt:2021yor} and $1S-3S$~\cite{Grinin:2020txk,Fleurbaey:2018fih} transitions in hydrogen in tension with the recently reported $2S-6P$ transition in hydrogen~\cite{Maisenbacher:2026nau}. 
While the Higgs portal interpretation is excluded by kaon decay constraints~\cite{NA62:2020xlg}, the featheron model remains phenomenologically viable.

We also present bounds on individual couplings to electrons, muons, and nucleons. 
For the latter, we consider both isosinglet ($g_p=g_n$) and isotriplet ($g_p=-g_n$) combinations. 
For most couplings, the data are consistent with the SM at the 
$3\sigma$ level. 
However, for purely electron-coupled interactions with sub-keV scalar mediator masses, a significant deviation is observed, driven primarily by the $2^3S-3^3D$ and $2^3P-3^3D$ transitions in $^4$He spectroscopy. 
See also the analysis in Ref.~\cite{Cong:2026kuv} based on the ionization energy of the $2^3S$ metastable state.
As discussed in Section~\ref{sec:fewelectrons}, the discrepancy between experimental measurements and theoretical predictions for these transitions is widely believed to originate from limitations in the current theoretical calculations rather than from new physics.
Assuming that this theoretical issue is resolved, the high experimental precision achieved in helium spectroscopy would make $^4$He the most sensitive probe of new electron-coupled interactions at very low-mass mediators in the sub-keV regime, surpassing constraints derived from the electron anomalous magnetic moment, $(g-2)_e$.

%%%
%%%

%%%%%%%%%%%%%%%%%%%%%%%%%%%%%%%%%%%%%%%%%%%%%%%%%%%%%%%%%%%%%%%%
\section{Summary and outlook}
\label{sec:Summary}
%%%%%%%%%%%%%%%%%%%%%%%%%%%%%%%%%%%%%%%%%%%%%%%%%%%%%%%%%%%%%%%%

In this review, we have surveyed a range of spectroscopic methods for probing physics beyond the Standard Model, with a particular emphasis on new force mediators with masses below the GeV scale. 
We discussed experimental approaches based on precision measurements of atomic and molecular transition frequencies and the associated energy shifts induced by new interactions. 
These techniques can be broadly divided into strategies relying on direct comparisons between theory and experiment, and those based on observables that are largely insensitive to detailed atomic-structure calculations. 
The latter exploit symmetries or special properties of atomic and molecular systems to enhance the relative impact of BSM effects compared to Standard Model contributions, thereby reducing theoretical uncertainties.

We also presented a general framework connecting ultraviolet descriptions of new physics, formulated at the level of quarks and leptons, to effective couplings of nucleons and nuclei that are directly probed in spectroscopic experiments. 
Within this framework, we summarized current state-of-the-art spectroscopic constraints on four representative benchmark models whose dominant signatures arise from spin-independent interactions, and on individual spin-independent couplings to electron, muon and nucleons.\\

Looking ahead, ongoing and future advances in spectroscopic techniques promise substantial gains in sensitivity. 
Rapid progress in optical clock technology, the development of highly charged ions, precision molecular spectroscopy, and emerging platforms such as nuclear clocks are expected to significantly extend the reach of spectroscopic searches. 
In parallel, improvements in atomic and nuclear theory, along with the construction of observables with reduced theoretical dependence, will play a crucial role in fully exploiting the experimental precision that is becoming available. 
Together, these developments position spectroscopy as an increasingly powerful and versatile tool for exploring new interactions.

A century ago, the effort to understand atomic structure led to the development of quantum mechanics and fundamentally reshaped our description of nature. 
Today, the extraordinary precision with which atomic and molecular systems can be measured and controlled offers a unique opportunity to probe physics beyond the Standard Model and may once again reveal new principles underlying fundamental interactions.

%%%%%%%%%%%%%%%%%%%%%%%%%%%%%%%%%%%%%%%%%%%%%%%%%%%%%%%%%%%%%%%%
\section*{ACKNOWLEDGMENTS}
%%%%%%%%%%%%%%%%%%%%%%%%%%%%%%%%%%%%%%%%%%%%%%%%%%%%%%%%%%%%%%%%

We thank Elina Fuchs, Teppei Kitahara, Fiona Kirk, Jeroen C.J. Koelemeij and Ben Ohayon for fruitful discussions and Jure Zupan for providing us with the data used in Table~\ref{tab:fhad}. 
CD is supported by the CNRS International Research Project NewSpec.
YS is supported by the ISF (grant No. 597/24) and by BSF (grant No. 2024091) and thanks CERN-TH for the scientific associateship. 

\bibliographystyle{ar-style5.bst}
\bibliography{references}

\end{document}